\documentclass[final,5p,times,twocolumn]{elsarticle}

\usepackage{amsmath}
\usepackage{amssymb}
\usepackage{graphicx}
\usepackage{booktabs}
\usepackage{multirow}
\usepackage{siunitx}
\usepackage{array}
\usepackage{tabularx}
\usepackage{algorithm}
\usepackage{algpseudocode}
\usepackage{enumitem}
\usepackage{url}
\usepackage{hyperref}
\usepackage{makecell}
\journal{Journal of Information Security and Applications}

\begin{document}

\begin{frontmatter}

\title{Behavioral Information Leakage in Darknet Traffic:
A Multi-Channel Analysis Across Anonymity Networks}

\author[aff1]{Javeriah Saleem\corref{cor1}}
\ead{jsaleem@csu.edu.au}

\author[aff2]{Rafiqul Islam}

\author[aff3,aff4]{Md Zahidul Islam}

\cortext[cor1]{Corresponding author}

\affiliation[aff1]{
    organization={School of Computing, Mathematics and Engineering,
    Charles Sturt University},
    city={Wagga Wagga},
    state={NSW},
    postcode={2650},
    country={Australia}
}

\affiliation[aff2]{
    organization={School of Computing, Mathematics and Engineering,
    Charles Sturt University},
    city={Albury},
    state={NSW},
    postcode={2640},
    country={Australia}
}

\affiliation[aff3]{
    organization={School of Computing, Mathematics and Engineering,
    Charles Sturt University},
    city={Bathurst},
    state={NSW},
    postcode={2795},
    country={Australia}
}

\affiliation[aff4]{
    organization={AI and Cyber Futures Center,
    Charles Sturt University},
    city={Bathurst},
    state={NSW},
    postcode={2795},
    country={Australia}
}

\begin{abstract}

Existing darknet traffic classification studies largely emphasize predictive accuracy while offering limited insight into the behavioral mechanisms that make encrypted services distinguishable. This paper proposes a behavioral information leakage framework that decomposes flow-level traffic into control, structural, and rhythmic descriptor groups across Tor, I2P, FreeNet, and ZeroNet. The framework combines normalized mutual-information analysis with Random-Forest-based predictive validation, structural--rhythmic interaction analysis, and cross-network service-variability evaluation under leakage-safe repeated stratified
cross-validation.

Results show that behavioral leakage varies considerably across anonymity networks. Tor achieves the highest service separability, with a Macro-F1 of 0.7165 and cumulative normalized leakage of 3.9461, whereas FreeNet exhibits the lowest combined leakage of 0.8744. Packet-size organization, directional exchange imbalance, packet tempo, and silence--burst behavior emerge as the main leakage mechanisms. The combined structural--rhythmic representation consistently provides the strongest within-network performance, while leave-one-network-out evaluation reveals limited transferability across anonymity architectures. The proposed Service Variability Index and Leakage Variability Index further show that video exhibits consistent network-specific separability, whereas chat and email demonstrate greater variability across anonymity-network pairs.

\end{abstract}


\end{frontmatter}


\section{Introduction}
\label{sec:introduction}

The increasing use of anonymity networks, such as Tor, I2P, FreeNet, and ZeroNet, has significantly transformed the landscape of private and encrypted online communication~\cite{saleem2024darknet}. These networks are designed to conceal user identities, communication endpoints, and browsing activities through layered routing, tunnel-based forwarding, and/or decentralized peer-to-peer communication infrastructures~\cite{saleem2022anonymity}. While these mechanisms improve privacy and resistance to surveillance, they also create challenges for network monitoring, digital forensics, malicious-activity detection, and network management. Consequently, darknet traffic analysis has become an important research area aimed at determining whether encrypted and anonymized traffic continues to expose distinguishable communication patterns at the flow level.

Existing darknet traffic classification studies have primarily focused on improving predictive performance through machine-learning and deep-learning techniques. A wide range of approaches have been proposed, including statistical feature engineering, convolutional architectures, recurrent neural networks, and hybrid learning frameworks~\cite{velan2015survey}. Although many of these studies report strong classification performance, they generally emphasize how accurately services can be classified rather than why different services remain identifiable despite anonymity protections. In most cases, traffic descriptors are treated as abstract predictive variables, with limited investigation into the behavioral mechanisms responsible for service distinguishability. As a result, the relationship between encrypted traffic behavior and information leakage remains insufficiently understood~\cite{velan2015survey,Montieri2020Dive}.

This limitation is particularly important in anonymity networks, where service identifiability may arise from multiple behavioral channels rather than isolated statistical features~\cite{Montieri2020Dive}. Some services may leak information through structural properties, such as encrypted payload-length structure, directional imbalance, and packet-size organization~\cite{Hayes2016KFingerprinting}, whereas others may leak information through rhythmic properties, such as inter-arrival timing, packet tempo, silence density, and burst behavior~\cite{panchenko2016website}. Existing studies rarely separate these behavioral dimensions explicitly, making it difficult to identify whether darknet services are primarily structural, rhythmic, or dependent on interactions between the two channels~\cite{RustNguyen2023Adversarial,Sharma2025Survey}. Furthermore, most prior work focuses either on individual networks or isolated classification settings, providing limited understanding of how behavioral leakage changes across different anonymity infrastructures and semantic service categories~\cite{RustNguyen2023Adversarial,Sharma2025Survey}.

To address these limitations, this paper proposes a behavioral leakage decomposition framework for darknet traffic analysis. Instead of treating traffic classification solely as a predictive task, the proposed framework determines how different categories of flow-level descriptors encode identifiable behavioral information. The flow-level descriptors are decomposed into three groups: control,
structural, and rhythmic. Structural descriptors characterize packet-size and payload-length organization, directional exchange, and subflow structure, whereas rhythmic descriptors capture temporal
dynamics, including inter-arrival variability, packet tempo, silence patterns, and burst irregularity. These representations are evaluated through global inter-network classification, intra-network service
analysis, and cross-network semantic comparison across Tor, I2P, FreeNet, and ZeroNet.

This study has five objectives. First, it investigates whether
encrypted darknet services exhibit measurable behavioral information leakage at the flow level. Second, it examines whether Structural or Rhythmic communication descriptors predominantly drive this leakage. Third, it evaluates whether combining structural and rhythmic descriptors provides complementary behavioral information beyond that
provided by either channel individually. Fourth, it identifies the descriptor groups and communication mechanisms responsible for service identifiability across different anonymity networks. Finally, it quantifies variation in network-specific service
separability and leakage magnitude using the proposed Service Variability Index (SVI) and Leakage Variability Index (LVI).

The main contributions of this paper, therefore, include:

\begin{itemize}[leftmargin=*]

    \item A behavioral information leakage framework for encrypted darknet traffic that decomposes flow-level descriptors into the control, structural, and rhythmic channels. This decomposition enables systematic analysis of the behavioral mechanisms underlying encrypted service identifiability across multiple anonymity networks.

    \item An integrated evaluation framework that combines information-theoretic leakage quantification with predictive validation, structural--rhythmic dominance, predictive synergy, and incremental cumulative leakage-gain analysis. These complementary perspectives characterize behavioral information exposure beyond conventional classification performance.

    \item A mechanism-level interpretation of behavioral leakage that links descriptor groups to observable communication behaviors. This analysis reveals how payload organization, imbalances in directional communication, packet tempo, and silence behavior contribute to service identifiability across Tor, I2P, FreeNet, and ZeroNet.

   \item A cross-network service-variability analysis based on the
proposed Service Variability Index (SVI) and Leakage Variability Index
(LVI). These indices quantify variation in network-specific service
separability and cumulative leakage across semantic network pairs.

\end{itemize}

The remainder of this paper is organized as follows. Section~\ref{sec:related-work} reviews relevant work on darknet traffic classification, encrypted traffic analysis, and behavioral interpretability. Section~\ref{sec:methodology} presents the proposed methodology, including the datasets used, the leakage-safe preprocessing protocol, the behavioral descriptor engineering, the experimental design and prediction validation procedures, and the evaluation metrics. Section~\ref{sec:results} presents the experimental results. These are reported in terms of cross-network behavioral transferability, the characteristics of descriptor-level leakage, the amount of behavioral leakage, and validated predictions. Additionally, this section includes a mechanism-level behavioral analysis, a structural--rhythmic interaction analysis, and a review of service-level variability of cross-network separability and leakage. Section 5 discusses the implications, limitations, and practical significance of the proposed framework. Finally, Section 6 concludes the paper and outlines the directions for future research.

\section{Background}
\label{sec:related-work}

The rapid growth of encrypted communication systems and anonymity-preserving overlay networks has fundamentally transformed modern network traffic analysis. The widespread adoption of end-to-end encryption, onion routing, tunnel-based forwarding, and decentralized peer-to-peer communication has significantly reduced the effectiveness of traditional payload-based inspection techniques. Consequently, cybersecurity research has progressively shifted from payload-centric traffic analysis techniques toward behavior-driven approaches. In this new paradigm, observable flow-level characteristics are used to infer communication behavior, application services, and network identities without decrypting a packet's contents. This transition has established encrypted traffic analysis as a critical research area for monitoring networks, gathering intelligence on cyberthreats, digital forensics, and detecting malicious activity.

Among all the anonymity-preserving infrastructures, Tor, I2P, FreeNet, and ZeroNet are among the most extensively studied darknet ecosystems. This is because they all have dual uses. On the one hand, they each provide essential privacy services, including anonymous communication, resistance to censorship, and the ability to share information securely for journalists, activists, and privacy-conscious users. However, this very guarantee of anonymity has also given rise to underground marketplaces, anonymous malware communication, illicit content distribution, and covert command-and-control channels. Consequently, determining whether encrypted darknet traffic continues to expose identifiable behavioral information despite strong cryptographic protection has become an important cybersecurity challenge~\cite{saleem2022anonymity}.

Although these anonymity networks pursue similar privacy objectives, they employ substantially different communication architectures. Tor relies on layered onion routing through volunteer-operated relay nodes, exposing residual characteristics such as packet-burst patterns, directional asymmetry, and timing irregularities. I2P employs independent inbound and outbound tunnels that introduce distinct latency distributions and directional communication dynamics. FreeNet and ZeroNet differ further by incorporating decentralized storage and peer-to-peer content distribution, resulting in communication behaviors that are fundamentally different from those of tunnel-based anonymity systems. These architectural differences suggest that identifying darknet services is unlikely to be possible by using a single statistical characteristic. Instead, the ability to differentiate different types of traffic emerges from multiple interacting behavioral mechanisms~\cite{saleem2024darknet}.

Existing research on encrypted darknet traffic analysis can be broadly categorized into four major directions: hierarchical traffic classification, machine-learning-based service identification, deep-learning-based encrypted traffic recognition, and robustness-oriented privacy studies~\cite{Shi2022WeightedKNN, Montieri2020Dive, RustNguyen2023Adversarial} . Collectively, these studies demonstrate that encrypted traffic is still distinguishable despite payload encryption. However, comparatively limited attention has been devoted to understanding why this is so.

Early studies on hierarchical classification, particularly those by Montieri et al~\cite{Montieri2020Dive}, demonstrate that statistical flow descriptors retain sufficient discriminative information to identify darknet traffic at multiple levels, including darknet detection, anonymity-network identification, and application-level service classification. These studies established the feasibility of hierarchical encrypted traffic recognition, demonstrating that flow-level statistical characteristics are informative even given encrypted communication. Nevertheless, the primary objective of these authors was to maximize hierarchical classification performance rather than investigate the behavioral mechanisms responsible for service distinguishability.

A second research direction focuses on supervised machine learning for darknet traffic classification. Numerous studies have employed statistical flow descriptors, including packet-length distributions, directional packet counts, byte statistics, flow duration, and inter-arrival timing, together with classifiers such as Random Forests, Support Vector Machines, decision trees, logistic regression, and gradient-boosting methods~\cite{Montieri2020Dive,Shi2022WeightedKNN,
Karunanayake2023ModifiedTor,RustNguyen2023Adversarial}. More recent work by Rust-Nguyen and Stamp~\cite{RustNguyen2023Adversarial} investigates adversarial robustness using both traditional machine-learning and deep-learning models. As such, they demonstrate that encrypted traffic classifiers are susceptible to carefully crafted perturbations. Although these studies confirm the effectiveness of statistical flow analysis, the descriptor sets were generally treated as undifferentiated predictive features rather than interpretable sources of behavioral information.

Deep learning has further advanced encrypted traffic analysis through convolutional neural networks (CNNs), recurrent neural networks (RNNs), long short-term memory (LSTM) networks, gated recurrent units (GRUs), and hybrid sequence-learning architectures that automatically learn complex traffic representations~\cite{Sirinam2018DeepFingerprinting,rimmer2017automated,
Diao2023ECGCN,Han2024DEGNN,Zou2024CADNet}. While these approaches have substantially improved predictive performance, they often sacrifice interpretability for accuracy. As highlighted by recent surveys~\cite{Luo2023BITization,Sharma2025Survey}, deep neural models generally provide limited insight into the observable behavioral characteristics responsible for service distinguishability, making it difficult to explain why encrypted applications remain identifiable even though their content is obfuscated ~\cite{Li2024Interaction}.

Another important line of research investigates the robustness and privacy implications of encrypted traffic fingerprinting. Studies on traffic morphing, packet padding, traffic splitting, timing perturbation, and adversarial obfuscation consistently demonstrate that classification performance is strongly influenced by modifications to observable communication behavior rather than the payload content itself~\cite{Juarez2016WTFPAD,Wang2017WalkieTalkie,
Cadena2020TrafficSliver, pulls2023maybenot}. Similarly, reports published by the European Union Agency for Cybersecurity (ENISA) emphasize that metadata, including packet timing, packet sizes, directional asymmetry, and flow dynamics, continue to expose exploitable behavioral side channels even under strong encryption~\cite{ENISA2020}. However, although these studies confirm that behavioral leakage exists, they generally evaluate robustness without systematically quantifying how different categories of behavioral descriptors differentiate among services.

Recent survey papers consistently identify several remaining challenges in encrypted traffic analysis, including limited interpretability, strong dataset dependence, weak cross-network generalization, and insufficient understanding of the behavioral mechanisms underlying encrypted traffic fingerprints~\cite{Hayes2016KFingerprinting,panchenko2016website,rimmer2017automated,Sirinam2018DeepFingerprinting}. Although model-explanation techniques, such as SHAP analysis and attention visualization, have been introduced, these approaches remain classifier-dependent and primarily identify influential predictors rather than quantifying behavioral information leakage itself~\cite{Lundberg2017SHAP}. Consequently, the relationships between how communications are structured, temporal dynamics, application semantics, and anonymity network architectures are still not well understood.

Motivated by these limitations, this work introduces an interpretability-driven behavioral leakage framework that extends existing encrypted traffic analysis beyond conventional predictive evaluation. Rather than treating statistical descriptors as a single feature space, encrypted traffic behavior is explicitly decomposed into control, structural, and rhythmic descriptor groups representing distinct communication mechanisms. Information-theoretic analysis is integrated with predictive
validation to quantify descriptor-level leakage, while
structural--rhythmic dominance, predictive synergy, and incremental
cumulative leakage gain characterize the contributions of the two behavioral channels. Finally, the proposed Service Variability Index (SVI) and Leakage Variability Index (LVI) quantify variation in network-specific separability and leakage magnitude across semantic
network pairs.

\section{Proposed Behavioral Leakage Framework}
\label{sec:methodology}

The proposed behavioral leakage decomposition framework evaluates the identifiability of darknet traffic across multiple anonymity networks. Unlike conventional approaches to darknet traffic classification, which primarily optimize prediction performance, the proposed framework investigates how different categories of flow-level descriptors expose behavioral information that remains observable despite payload encryption and anonymity-network routing. The framework combines a descriptor taxonomy with analytical
procedures that quantify information-theoretic leakage, validate predictions, analyze dominance, evaluate behavioral synergies, and interpret descriptor-level information to identify the mechanisms that underpin darknet service distinguishability. Figure~\ref{fig:framework} presents the overall research framework, while Algorithm 1 summarizes the full analysis procedure.

\subsection{Dataset}
\label{subsec:dataset}

The experiments in this study were conducted using the Darknet Dataset 2020, a publicly available flow-level darknet traffic dataset containing encrypted communication traces collected from multiple anonymity-network environments~\cite{Hu2020DarknetDataset}. The dataset includes traffic found in today's major anonymity infrastructures -- Tor, I2P, FreeNet, and ZeroNet -- together with their corresponding application categories at the service level. Each traffic flow is represented using statistical descriptors extracted from encrypted packet streams to allow traffic analysis without needing to inspect the payload.

After data cleaning and removing incomplete records, the final dataset contained \(48{,}644\) traffic flows distributed across the four anonymity networks. The data span multiple service categories including browsing, chat, email, file transfer, peer-to-peer communications, audio, video, and VoIP. The sheer diversity of the traffic means we could test the framework under heterogeneous conditions but with semantically similar services operating on different anonymity infrastructures.

Unlike the packet-sequence datasets commonly used in deep-learning approaches, the Darknet Dataset 2020 provides flow-level statistical representations, making it particularly suitable for descriptor-level interpretations and behavioral decomposition analysis. The original traffic features describe observable communication characteristics, such as packet-size distributions, directional traffic exchange, flow duration, timing variability, inter-arrival behavior, activity patterns, and subflow characteristics. These statistical properties preserve the behavioral traces generated by service operation even when an application's payloads are encrypted.

To use the framework, we reorganized the original feature space  into three complementary descriptor groups, each representing a distinct source of behavioral information leakage. The control group (\(C\)), consisting of six descriptors, contained protocol-level and metadata-related information. The structural group (\(S\)) captured packet-size distributions, directional communication patterns, and traffic-exchange characteristics through 15 descriptors. The rhythmic group (\(R\)) representing temporal communication behavior included packet tempos, burst dynamics, silence patterns, and timing variability spanning 15 descriptors. We then constructed a further combined structural--rhythmic representation (\(S+R\)) by integrating all structural and rhythmic descriptors to yield a 30-descriptor behavioral representation.

To comprehensively evaluate behavioral information leakage under different classification conditions, the dataset was reorganized into three complementary experimental settings:
\begin{figure*}[!t]
    \centering
    \includegraphics[
        width=0.6\textwidth,
        trim=3 3 3 3,
        clip
    ]{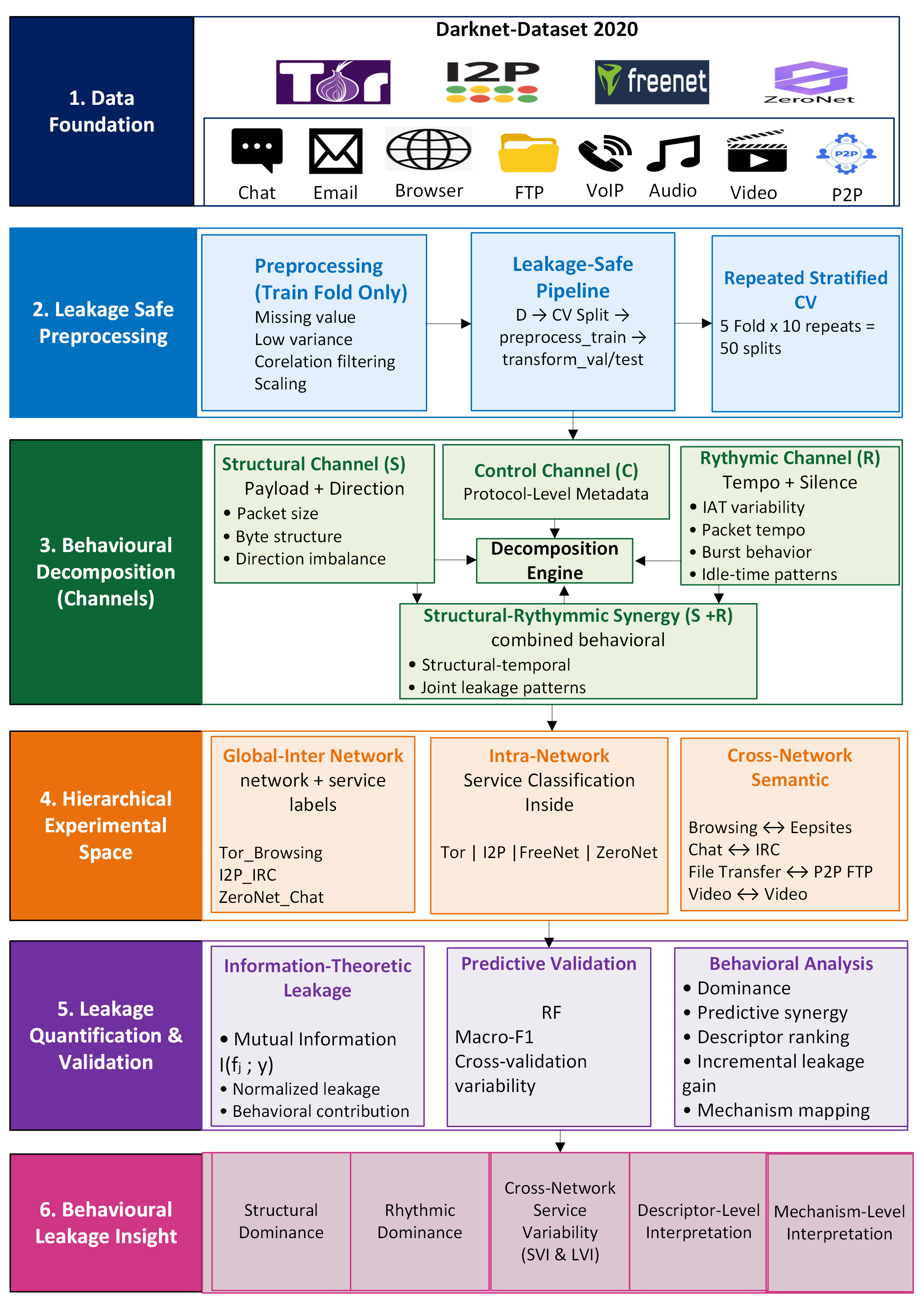}
    \caption{Proposed behavioral information leakage analysis framework.}
    \label{fig:framework}
\end{figure*}
\begin{algorithm}[!t]
\caption{Behavioral Information Leakage Analysis Framework}
\label{alg:behavioral-leakage}
\small

\begin{algorithmic}[1]

\Require Flow-level darknet dataset \(\mathcal{D}\)
\Require Descriptor groups \(\mathcal{G}=\{C,S,R,S+R\}\)

\Ensure Descriptor-level, mechanism-level, and service-level behavioral leakage profiles

\State Clean \(\mathcal{D}\) and remove invalid or incomplete samples
\State Construct global, intra-network, and cross-network semantic label spaces

\State Define the descriptor groups:
\Statex \hspace{\algorithmicindent}
\(C \gets\) Control descriptors
\Statex \hspace{\algorithmicindent}
\(S \gets\) Structural descriptors
\Statex \hspace{\algorithmicindent}
\(R \gets\) Rhythmic descriptors
\Statex \hspace{\algorithmicindent}
\(S+R \gets S \cup R\)

\ForAll{experimental settings \(e \in \mathcal{E}\)}
    \ForAll{descriptor groups \(g \in \mathcal{G}\)}
        \ForAll{repeated stratified cross-validation splits}

            \State Partition the data into
            \(\mathcal{D}_{\mathrm{train}}\) and
            \(\mathcal{D}_{\mathrm{val}}\)

            \State Fit low-variance filtering, correlation filtering, imputation, and robust scaling using \(D_{\mathrm{train}}\)

            \State Apply the fitted transformations to
            \(\mathcal{D}_{\mathrm{val}}\)

            \State Estimate descriptor-level mutual information \(I(f_j;Y)\)

            \State Obtain normalized cumulative leakage \(L_g\)

            \State Train a Random Forest classifier using descriptor group \(g\)

            \State Evaluate Macro-F1 and record its variability across repeated splits

        \EndFor

        \State Aggregate leakage and predictive results across all splits

    \EndFor
\EndFor

\State Rank descriptors according to normalized mutual information
\State Map the ranked descriptors to behavioral communication mechanisms

\State Calculate interaction and cross-network variability measures:

\Statex \hspace{\algorithmicindent}
\(\displaystyle
D_{\mathrm{RF}} = F_{1}^{S}-F_{1}^{R}
\)

\Statex \hspace{\algorithmicindent}
\(\displaystyle
D_{\mathrm{MI}} = L_S-L_R
\)

\Statex \hspace{\algorithmicindent}
\(\displaystyle
\Delta_{\mathrm{RF}}
=
F_{1}^{S+R}
-
\max(F_{1}^{S},F_{1}^{R})
\)

\Statex \hspace{\algorithmicindent}
\(\displaystyle
\Delta_{L}
=
L_{S+R}-\max(L_S,L_R)
\)

\ForAll{services evaluated across semantic network pairs}

    \Statex \hspace{\algorithmicindent}
    \(\displaystyle
    \mathrm{SVI}
    =
    \sigma\!\left(
    F_1^{(1)},F_1^{(2)},\ldots,F_1^{(n)}
    \right)
    \)

    \Statex \hspace{\algorithmicindent}
    \(\displaystyle
    \mathrm{LVI}
    =
    \sigma\!\left(
    L_1,L_2,\ldots,L_n
    \right)
    \)

\EndFor

\State Generate descriptor-level, mechanism-level, and service-level behavioral leakage profiles

\end{algorithmic}
\end{algorithm}
\begin{itemize}[leftmargin=*]

    \item \textbf{Global inter-network setting:} Traffic from all anonymity networks and services was combined into a unified classification space, enabling evaluation of behavioral leakage across the entire darknet ecosystem.

    \item \textbf{Intra-network setting:} Service-level classification was performed independently within each anonymity network, namely Tor, I2P, FreeNet, and ZeroNet, to quantify network-specific behavioral leakage mechanisms.

    \item \textbf{Cross-network semantic setting:} Semantically equivalent services operating across different anonymity networks were compared to quantify network-origin separability and variation in cumulative behavioral leakage.

\end{itemize}

This experiment design meant the framework could not only analyze whether the darknet services were identifiable but also how behavioral leakage varies across networks, services, communication architectures, and semantic application categories. Consequently, the dataset provided a suitable foundation for decomposing service identifiability into the control, structural, and rhythmic channels and for investigating their relative contributions across multiple anonymity network environments.

\subsection{Leakage-Safe Preprocessing and Validation Protocol}
\label{subsec:preprocessing}

A primary objective of this study is to ensure that behavioral leakage measurements reflect genuine service identification signals rather than artifacts introduced by data preprocessing. To prevent information leakage between the training and validation data, all preprocessing operations were performed exclusively on the training portion of each validation split and subsequently applied to the corresponding validation data using the fitted transformation parameters.

The preprocessing pipeline consisted of four sequential stages.

\begin{enumerate}
    \item \textbf{Low-variance filtering:} Numeric descriptors with
    training-fold variance less than or equal to
    \(\tau_v=1\times10^{-10}\) were removed.

    \item \textbf{Correlation filtering:} Redundant numeric
    descriptors were filtered using an absolute Pearson-correlation
    threshold of \(\tau_{\rho}=0.95\). When a descriptor exhibited
    \(\lvert\rho\rvert\geq0.95\) with any earlier descriptor in the
    existing feature order, the later-appearing descriptor was removed.

    \item \textbf{Missing-value imputation:} After feature filtering,
    missing numerical values were replaced using the median computed
    from the training fold. The categorical protocol descriptor, when
    present, was imputed using its most frequent training-fold value.

    \item \textbf{Robust feature scaling:} Numerical descriptors were
    scaled using the training-fold median and interquartile range.
    The categorical protocol descriptor was one-hot encoded, with
    previously unseen validation categories ignored.
\end{enumerate}

All preprocessing parameters were fixed before model evaluation. Numeric descriptors with training-fold variance less than or equal to \(\tau_v=1\times10^{-10}\) were removed. Redundant numeric descriptors were subsequently filtered using an absolute Pearson correlation threshold of \(\tau_{\rho}=0.95\). The upper triangular portion of the training-fold correlation matrix was examined according to the existing descriptor order. When a descriptor exhibited \(\lvert\rho\rvert \geq 0.95\) with any earlier descriptor, the later-appearing descriptor was removed and the earlier descriptor was retained. Non-numeric descriptors were preserved and processed separately. Both variance and correlation filtering were fitted independently within each training fold, and the retained descriptor set was then applied unchanged to the corresponding validation fold.

Descriptor-level mutual information was estimated independently within each training fold after low-variance filtering, correlation filtering, imputation, and robust scaling. The resulting descriptor-level values were normalized by label entropy and aggregated within each descriptor group to obtain cumulative behavioral leakage. The identical fitted
preprocessing transformation was applied to the corresponding validation fold, preventing information from the validation data from influencing descriptor selection or leakage estimation.

This train-fold-only protocol ensured that descriptor engineering, mutual information estimation, and predictive validation remained free from train--test contamination. Consequently, the reported behavioral leakage measurements only reflect the information genuinely available to a classifier, rather than any information inadvertently introduced during preprocessing.

\subsection{Behavioral Descriptor Engineering}
\label{subsec:descriptor-engineering}

The central premise of this study is that identifying services within anonymity networks is a function of recognizing multiple behavioral mechanisms rather than a single source of information. To systematically investigate these mechanisms, we decomposed the original traffic representation into three complementary descriptor groups corresponding to distinct behavioral leakage channels. Let the behavioral representation of the \(i\)-th traffic flow be defined as

\begin{equation}
\mathbf{B}_{i}=\left\{C_{i},S_{i},R_{i}\right\},
\label{eq:behavior_representation}
\end{equation}

where \(C_{i}\) denotes the Control descriptor group, \(S_{i}\) represents the Structural descriptor group, and \(R_{i}\) represents the Rhythmic descriptor group.

\subsubsection{Control Descriptor Group (\texorpdfstring{$C$}{C})}

The control descriptor group provides baseline communication information and is not intended to directly characterize behavioral communication patterns. Typical control descriptors include protocol identifiers, flow-control statistics, and directional metadata that may indirectly contribute to traffic separability.

The control descriptor group serves two primary purposes. First, it provides a baseline reference for evaluating behavioral leakage independently of protocol-level information. Second, it enables direct comparison between traditional traffic descriptors and behavior-oriented descriptor groups. Although control descriptors may improve predictive performance, they generally provide limited insight into the underlying communication behavior responsible for distinguishing encrypted darknet services.

\subsubsection{Structural Descriptor Group (\texorpdfstring{$S$}{S})}

The structural descriptor group characterizes how information is spatially organized within encrypted traffic flows. These descriptors capture observable communication structures without requiring access to the encrypted payload content. Structural behavior primarily reflects how services exchange information across encrypted channels and how communication volumes are distributed between the forward and backward traffic directions.

The structural descriptor group contains descriptors associated with packet-size distributions, packet-length variability, payload-per-packet behavior, directional packet imbalance, forward/backward byte asymmetry, subflow organization, and packet aggregation characteristics. The descriptor set is defined as

\begin{equation}
S=\left\{s_{1},s_{2},\ldots,s_{m}\right\},
\label{eq:structural_set}
\end{equation}

where \(m\) denotes the total number of structural descriptors.

Unlike temporal descriptors, structural descriptors are relatively insensitive to communication timing. Instead, they characterize how communication volumes and directional exchanges are organized within each traffic flow. Consequently, structural behavior often reflects application-level communication semantics, including browsing intensity, file-transfer organization, streaming behavior, and peer-to-peer communication patterns.

\subsubsection{Rhythmic Descriptor Group (\texorpdfstring{$R$}{R})}

The rhythmic descriptor group characterizes temporal communication behavior within encrypted traffic flows. Rather than describing communication structures, rhythmic descriptors capture when and how frequently communication events occur. These descriptors model the temporal dynamics of encrypted traffic and frequently remain observable despite payload encryption and anonymity protection.

The rhythmic descriptor group includes descriptors associated with inter-arrival time variability, packet tempo, silence density, idle behavior, burst irregularity, directional timing asymmetry, and communication pacing characteristics. The descriptor space is defined as

\begin{equation}
R=\left\{r_{1},r_{2},\ldots,r_{n}\right\},
\label{eq:rhythmic_set}
\end{equation}

where \(n\) denotes the total number of rhythmic descriptors.

Rhythmic descriptors are particularly important because many anonymity-network services exhibit distinct communication rhythms even when packet contents remain encrypted. Interactive applications such as browsing and chat typically generate highly irregular timing behavior, whereas streaming and bulk-transfer services generally exhibit more stable temporal communication patterns.

\subsubsection{Combined Structural--Rhythmic Representation (\texorpdfstring{$S+R$}{S+R})}

Although structural and rhythmic descriptors independently capture different behavioral characteristics, identifying encrypted services can also be done from the interactions between these two complementary channels. Evaluating this hypothesis involved combining the structural and rhythmic descriptor groups into a unified behavioral representation defined as

\begin{equation}
\mathbf{B}_{i}^{(S+R)} = S_{i}\cup R_{i},
\label{eq:combined_representation}
\end{equation}

where \(\cup\) denotes the union of the structural and rhythmic descriptor sets.

The combined representation enables evaluation of whether Structural and Rhythmic descriptors provide complementary predictive information. An improvement in Macro-F1 beyond the stronger individual descriptor group provides evidence of predictive complementarity. An increase in cumulative leakage indicates additional aggregate descriptor-level exposure but is interpreted separately because the combined representation contains more descriptors and may include overlapping marginal mutual-information contributions.

The proposed behavioral decomposition therefore transforms darknet traffic analysis from a conventional feature-engineering problem into an interpretability-driven behavioral leakage analysis framework, where encrypted communication is analyzed through structural organization, temporal rhythm, and the interactions between these complementary behavioral mechanisms across multiple anonymity networks.

\subsection{Experimental Design}
\label{subsec:experimental-design}

To comprehensively evaluate behavioral information leakage across anonymity networks, the proposed framework was designed around three complementary experimental settings: global inter-network classification, intra-network service analysis, and cross-network semantic comparison. These settings were selected to analyze the identifiability of darknet traffic under progressively varying behavioral conditions while isolating the influence of network architecture, service semantics, and descriptor-group interactions.

Unlike conventional darknet traffic classification studies that evaluate only a single classification scenario, the proposed framework investigates how structural and rhythmic behavioral leakage varies across multiple operational contexts. This design enabled systematic evaluation of network-specific
separability, service-level leakage variation, and
architecture-dependent Structural or Rhythmic dominance across the evaluated anonymity networks.

The complete experimental design is represented as

\begin{equation}
\mathcal{E}
=
\left\{
E_{\mathrm{global}},
E_{\mathrm{intra}},
E_{\mathrm{semantic}}
\right\},
\label{eq:experiment-space}
\end{equation}

where \(E_{\mathrm{global}}\), \(E_{\mathrm{intra}}\), and
\(E_{\mathrm{semantic}}\) denote the global inter-network,
intra-network, and cross-network semantic experiments,
respectively.

For every experimental setting
\(e\in\mathcal{E}\),
we evaluated four independent descriptor configurations:

\begin{equation}
\mathcal{G}
=
\left\{
C,
S,
R,
S+R
\right\},
\label{eq:descriptor-groups}
\end{equation}

where \(C\), \(S\), \(R\), and \(S+R\) represent the control,
structural, rhythmic, and combined structural--rhythmic descriptor
groups, respectively.

This design enabled us to directly compare the descriptor
groups under identical classification conditions.


\subsubsection{Global Inter-Network Classification}

The first experimental setting was designed to evaluate the distinguishability of
darknet services across all anonymity networks simultaneously. In
this setting, traffic from Tor, I2P, FreeNet, and ZeroNet was merged
into a unified classification space defined as

\begin{equation}
y_{\mathrm{global}}
=
\left\{
\text{network}
+
\text{service}
\right\},
\label{eq:global-label-space}
\end{equation}

where each class corresponds to a network-specific service identity,
for example, \textit{Tor\_Browsing},
\textit{I2P\_IRC},
\textit{FreeNet\_Video}, and
\textit{ZeroNet\_Chat}.

In this setting, we evaluated the overall behavioral separability of darknet
services across different anonymity infrastructures. Since both
the network architecture and the service semantics coexist within the same
classification problem, we used global inter-network classification to capture the variations in
both inter-networks and intra-service behaviors.
Consequently, this setting closely reflects practical deployment
scenarios in which encrypted traffic originating from multiple
anonymity systems coexists within a common monitoring environment.


\subsubsection{Intra-Network Service Classification}

The second experimental setting involved independently evaluating service distinguishability
 within each anonymity network. For every network
\(n\in\mathcal{N}\), service-level classification was defined as

\begin{equation}
y_{\mathrm{intra}}^{(n)}
=
\left\{
\text{service}
\mid
\text{network}=n
\right\},
\label{eq:intra-label-space}
\end{equation}

where

\[
\mathcal{N}
=
\{
\text{Tor},
\text{I2P},
\text{FreeNet},
\text{ZeroNet}
\}.
\]

This experimental setting removes cross-network architectural
variation and isolates behavioral leakage originating solely from
service-level communication behavior. Accordingly, we conducted four independent
classification experiments, one each to classify services in Tor, I2P,
FreeNet, and ZeroNet.

The intra-network setting enabled us to directly evaluate whether
services remain behaviorally distinguishable when all traffic belongs
to the same anonymity infrastructure. Consequently, differences
observed among descriptor groups can be interpreted as evidence of
service-specific behavioral leakage rather than network-level
architectural separation.


\subsubsection{Cross-Network Semantic Comparison}

The third experimental setting was designed to quantify how strongly
semantically equivalent services reveal their anonymity-network
origin. Rather than comparing unrelated applications, each experiment
contains traffic from the same semantic service operating over two
different anonymity infrastructures.

For each semantic service category,

\begin{equation}
E_{\mathrm{semantic}}^{(s)}
=
\left\{
\mathrm{service}_{s}^{(n_1)},
\mathrm{service}_{s}^{(n_2)}
\right\},
\end{equation}

where \(s\) denotes a shared semantic service and \(n_1\) and \(n_2\)
represent different anonymity networks.

Examples include:
\begin{itemize}
    \item Tor Browsing versus I2P Eepsites;
    \item Tor Chat versus I2P IRC;
    \item Tor FTP versus I2P FTP; and
    \item Tor Video versus FreeNet Video.
\end{itemize}

In each pairwise experiment, the class label represents the anonymity
network rather than the service category. High Macro-F1 therefore
indicates that the same semantic service retains strong
architecture-specific differences, whereas lower Macro-F1 indicates
greater similarity between the two network implementations. A total of
31 cross-network semantic experiments were performed across browsing,
chat, email, FTP, P2P, audio, and video services.

\subsubsection{Cross-Network Transferability Evaluation}
\label{subsubsec:cross_network_transfer}

In addition to the three primary classification settings, two
complementary transferability evaluations were conducted to determine
whether behavioral representations learned from one anonymity-network
environment generalize to another.

First, pairwise cross-network transferability was evaluated using the
combined structural--rhythmic representation. For each source--target
network pair, the model was trained using traffic from one source
network and evaluated exclusively using traffic from a different
target network. This evaluation measures the extent to which service-level behavioral patterns learned from one anonymity infrastructure remain discriminative under another infrastructure.

For each source--target pair, the evaluation was restricted to service labels present in both networks. Source--target combinations containing fewer than two shared service categories were excluded. Consequently, the number and identity of evaluated service classes could differ across network pairs, and the resulting pairwise Macro-F1 values were interpreted within their corresponding shared-service spaces.

Second, a leave-one-network-out evaluation was conducted. In each iteration, one anonymity network was held out for testing, while traffic from the remaining three networks was pooled for training. The Structural, Rhythmic, and combined structural--rhythmic representations were evaluated separately. Macro-F1 was used to measure transferability because the service categories were imbalanced.

For leave-one-network-out evaluation, the class space was restricted to service categories present in both the held-out network and the pooled training networks. Traffic belonging to unmatched service categories was excluded before model training and testing. Each network-defined transfer evaluation used the complete eligible training and test partitions and was not performed through repeated random cross-validation.

For both transferability protocols, all preprocessing operations, including low-variance filtering, correlation filtering, imputation, and robust scaling, were fitted using the training data only.

\subsubsection{Cross-Validation Strategy}
\label{subsubsec:cv_strategy}

To ensure statistically reliable evaluation under heterogeneous service distributions and class imbalance, the global, intra-network, and semantic-pair experiments employed repeated stratified cross-validation. Stratification preserves class distributions across the training and validation partitions while reducing the variance introduced by random partitioning.

For each split,

\begin{equation}
\mathcal{D}
=
\mathcal{D}_{\mathrm{train}}
\cup
\mathcal{D}_{\mathrm{val}},
\label{eq:dataset-split}
\end{equation}

subject to

\begin{equation}
\mathcal{D}_{\mathrm{train}}
\cap
\mathcal{D}_{\mathrm{val}}
=
\varnothing,
\label{eq:dataset-disjoint}
\end{equation}

where
\(\varnothing\)
denotes the empty set.

Repeated stratified five-fold cross-validation was performed ten times
using \texttt{random\_state=42}, yielding 50 matched
training--validation splits for each global, intra-network, and
semantic-pair experimental configuration. The same splits were used
across the descriptor groups to support paired comparison of their
predictive and leakage results.

All preprocessing operations, descriptor filtering, robust scaling,
behavioral leakage estimation, and predictive validation were
performed independently within each training partition to prevent
information leakage from the validation data.

Repeated evaluation provided estimates of predictive separability and descriptor-level behavioral leakage, together with their variability across the 50 matched splits.

\subsection{Predictive Validation and Evaluation Metrics}
\label{subsec:evaluation}

The proposed behavioral decomposition framework both quantifies information-theoretic leakage and validates predictions to provide a comprehensive evaluation of encrypted darknet traffic. While mutual information is used to estimate the amount of behavioral information contained within different descriptor groups, predictive validation determines whether the exposed information is sufficient to reliably  discriminate between services. Rather than proposing a new classification algorithm, the predictive model serves as an independent validation mechanism for assessing the practical utility of the quantified behavioral leakage. Consequently, the proposed framework evaluates each descriptor group from predictive, information-theoretic, interaction, and cross-network service-variability perspectives.

Random Forest was selected because bootstrap aggregation and randomized
feature selection reduce the variance of individual decision trees,
while the ensemble can capture nonlinear relationships and interactions
among heterogeneous flow-level descriptors ~\cite{Breiman2001RF}.
The Random Forest classifier used 300 trees with Gini impurity,
bootstrap sampling, and square-root feature sampling at each split.
Tree depth was unrestricted, with
\texttt{min\_samples\_split=2} and
\texttt{min\_samples\_leaf=1}. To account for class imbalance,
class weights were recomputed for each bootstrap sample using
\texttt{class\_weight=balanced\_subsample}. All experiments used
\texttt{random\_state=42} and parallel execution with
\texttt{n\_jobs=-1}.
To ensure unbiased evaluation, the protocol for validating the predictions is leakage-safe, as described in Section~\ref{subsec:preprocessing}. Every iteration of the stratified cross-validation includes the following steps: preprocessing, descriptor engineering, and mutual information estimation. These are performed exclusively on the training partition before the learned transformations are applied to the corresponding validation partition. This procedure prevents train--validation information leakage and ensures a fair comparison across all descriptor groups and experimental settings.

For every experimental configuration, we trained four independent Random Forest classifiers; one for each of the control, structural, rhythmic, and combined structural--rhythmic descriptor groups. The resulting predictions were evaluated using complementary predictive and information-theoretic performance measures.

\subsubsection{Predictive Performance}

Predictive performance refers to the discriminative capability of each descriptor representation across repeated stratified cross-validation folds. Since the dataset contains imbalanced classes across multiple service categories, we preferred metrics that assign equal importance to every class as opposed to overall classification accuracy.

We therefore chose the Macro-F1 score as the primary metric for measuring predictive performance. Macro-F1 is defined as the arithmetic mean of the class-wise F1 scores~\cite{Sokolova2009Metrics},

\begin{equation}
\mathrm{Macro}\text{-}F_{1}
=
\frac{1}{K}
\sum_{k=1}^{K}
F_{1}^{(k)},
\label{eq:macrof1}
\end{equation}

where \(K\) denotes the total number of service classes.

In addition to mean predictive performance, the standard deviation
of Macro-F1 across the 50 repeated stratified splits is reported to
characterize sensitivity to data partitioning. Cross-network variability in network-specific service separability is
evaluated separately using the proposed Service Variability Index
(SVI), formally defined in
Section~\ref{subsubsec:stability_analysis}.


\subsubsection{Behavioral Information Leakage}

Predictive performance alone cannot quantify the amount of behavioral information exposed by encrypted traffic~\cite{Kraskov2004MI,Ross2014MI}. Therefore, we quantified the amount of  information exposure at the descriptor level using the mutual information between descriptor \(f_j\) and service label \(Y\),

\begin{equation}
I(f_j;Y)
=
\sum_{f_j}
\sum_{Y}
p(f_j,Y)
\log
\left(
\frac{p(f_j,Y)}
{p(f_j)\,p(Y)}
\right),
\label{eq:mutual_information}
\end{equation}

where \(p(f_j,Y)\) denotes the joint probability distribution and \(p(f_j)\) and \(p(Y)\) denote the corresponding marginal probability distributions.

Most engineered flow descriptors are continuous. The Control representation additionally contains a categorical protocol descriptor, which was imputed using its most frequent training-fold value and subsequently one-hot encoded. During mutual-information
estimation, the resulting binary protocol columns were marked as discrete, while the remaining numerical descriptors were treated as continuous.

Descriptor-level mutual information was estimated using
\texttt{mutual\_info\_classif} with
\texttt{n\_neighbors=3} and
\texttt{random\_state=42}. Service labels were encoded before estimation, and the estimator was fitted independently within each training fold following leakage-safe preprocessing. Label entropy was
calculated using a base-2 logarithm, and cumulative normalized leakage was obtained by dividing the sum of descriptor-level mutual-information values by \(H(Y)\).

The cumulative normalized leakage associated with descriptor group
\(g\) is defined as

\begin{equation}
L_g
=
\frac{1}{H(Y)}
\sum_{f_j \in g}
I(f_j;Y),
\label{eq:cumulative_leakage}
\end{equation}

where \(H(Y)\) denotes the entropy of the service label and
\(I(f_j;Y)\) denotes the marginal mutual information between
descriptor \(f_j\) and label \(Y\). The resulting quantity represents
an aggregate descriptor-level leakage score rather than the joint
mutual information of the complete descriptor group. Consequently,
\(L_g\) may contain overlapping descriptor contributions and is not
restricted to the interval \([0,1]\).

Because cumulative leakage depends partly on the number of retained
descriptors, comparisons involving the combined \(S+R\)
representation are interpreted together with predictive performance
and descriptor-level leakage distributions rather than as standalone
evidence of information-theoretic synergy.


\subsubsection{Structural--Rhythmic Interaction Analysis}
\label{subsubsec:interaction_analysis}

To determine how the Structural and Rhythmic descriptor groups
contribute to behavioral information leakage, their relative
dominance and complementary predictive contribution were evaluated
from predictive and information-theoretic perspectives.

Predictive dominance is defined as

\begin{equation}
D_{\mathrm{RF}}
=
F_{1}^{S}
-
F_{1}^{R},
\label{eq:rf_dominance}
\end{equation}

where \(F_{1}^{S}\) and \(F_{1}^{R}\) denote the Macro-F1 scores
obtained using the Structural and Rhythmic descriptor groups,
respectively. Information-theoretic dominance is defined as

\begin{equation}
D_{\mathrm{MI}}
=
L_{S}
-
L_{R},
\label{eq:mi_dominance}
\end{equation}

where \(L_{S}\) and \(L_{R}\) represent the corresponding cumulative
normalized leakage values. Positive values indicate Structural
dominance, negative values indicate Rhythmic dominance, and values
close to zero indicate no clear dominance.

The complementary predictive contribution of the two descriptor
groups is quantified as

\begin{equation}
\Delta_{\mathrm{RF}}
=
F_{1}^{S+R}
-
\max
\left(
F_{1}^{S},
F_{1}^{R}
\right),
\label{eq:rf_synergy}
\end{equation}

where \(F_{1}^{S+R}\) denotes the Macro-F1 obtained using the combined
structural--rhythmic representation. A positive
\(\Delta_{\mathrm{RF}}\) indicates that combining the two behavioral
channels improves predictive separability beyond the stronger
individual channel.

The corresponding incremental cumulative leakage gain is defined as

\begin{equation}
\Delta_{L}
=
L_{S+R}
-
\max
\left(
L_{S},
L_{R}
\right),
\label{eq:incremental_leakage_gain}
\end{equation}

where \(L_{S+R}\) denotes cumulative normalized leakage for the
combined representation. The quantity \(\Delta_L\) measures the
additional aggregate descriptor-level leakage exposed by \(S+R\)
beyond the stronger individual group. It is not interpreted as formal
information-theoretic synergy because cumulative leakage is obtained
by summing marginal descriptor-level mutual-information
contributions.


\subsubsection{Cross-Network Service Variability Analysis}
\label{subsubsec:stability_analysis}

Dominance and synergy characterize interactions between descriptor
groups but do not indicate whether the degree of network-specific
separability remains consistent for the same semantic service across
different anonymity-network pairs. This property was evaluated using
the Service Variability Index (SVI) and Leakage Variability Index
(LVI).

The Service Variability Index is defined as

\begin{equation}
\mathrm{SVI}
=
\sigma
\left(
F_{1}^{(1)},
F_{1}^{(2)},
\ldots,
F_{1}^{(n)}
\right),
\label{eq:svi}
\end{equation}

where \(F_{1}^{(i)}\) denotes the Macro-F1 obtained when
distinguishing the anonymity-network origin of a fixed semantic
service in the \(i\)-th network-pair comparison. A low SVI indicates
that the degree of network-specific separability remains consistent
across the evaluated network pairs. It does not imply that the service
has an architecture-invariant traffic signature.

The Leakage Variability Index is defined as

\begin{equation}
\mathrm{LVI}
=
\sigma
\left(
L_1,
L_2,
\ldots,
L_n
\right),
\label{eq:lvi}
\end{equation}

where \(L_i\) represents the cumulative normalized leakage associated
with the corresponding semantic network-pair comparison. A low LVI
indicates that the magnitude of exposed behavioral information remains
consistent across network pairs.

Together, SVI and LVI quantify the variability of network-specific service separability and leakage magnitude. They are descriptive cross-network variability measures rather than direct measures of service-label transferability.

For descriptive visualization, the observed service profiles were partitioned using operational cut-offs of
\(\tau_{\mathrm{SVI}}=0.020\) and \(\tau_{\mathrm{LVI}}=1.500\). These cut-offs were used only to summarize the profiles shown in Figure~\ref{fig:service_variability} and are not proposed as universal cross-network variability thresholds.

\section{Results}
\label{sec:results}

This section evaluates behavioral information leakage across anonymity networks using the proposed behavioral decomposition framework. The results are presented according to the analytical workflow illustrated in Figure~\ref{fig:framework}. Rather than treating darknet traffic classification solely as a prediction task, the analysis investigates how behavioral information propagates through the structural and rhythmic communication channels and how these channels contribute to service identifiability across different anonymity networks.

The experimental evaluation comprised six stages. First, we examined cross-network transferability to determine whether behavioral signatures generalize across anonymity network infrastructures. Second, we conducted a descriptor-level analysis to identify the specific behavioral variables responsible for information leakage. Third, we estimated and compared the information-theoretic leakage with the results of validating the predictions to determine whether machine-learning models can effectively exploit the exposed behavioral information. Fourth, we decomposed behavioral leakage into interpretable communication mechanisms to identify the underlying sources of service distinguishability. Fifth, we analyzed the Structural and Rhythmic descriptor groups
both independently and jointly to quantify dominance and synergy effects. Finally, we investigated how network-origin separability and leakage magnitude varied across semantic service categories.

Collectively, these analyses provided a comprehensive characterization of behavioral information leakage in encrypted darknet traffic. Moreover, they revealed how different anonymity networks expose distinct leakage mechanisms despite operating under similar privacy-preserving objectives.
\begin{table*}[!htbp]
\centering
\caption{Leave-one-network-out cross-network transferability results.}
\label{tab:lono}

\small
\setlength{\tabcolsep}{12pt}
\renewcommand{\arraystretch}{1.2}

\begin{tabular}{lcccccl}
\toprule
\textbf{Held-out Network} &
\textbf{Structural} &
\textbf{Rhythmic} &
\textbf{Combined} &
\(\boldsymbol{S-R}\) &
\textbf{Fusion Gain} &
\textbf{Relative Channel Pattern} \\
\midrule
FreeNet & 0.1667 & 0.1618 & 0.1657 & 0.0049  & -0.0010 & Near-balanced \\
I2P     & 0.1716 & 0.1647 & 0.1591 & 0.0069  & -0.0125 & Near-balanced \\
Tor     & 0.1223 & 0.1822 & 0.1175 & -0.0600 & -0.0647 & Rhythmic-favoured \\
ZeroNet & 0.0908 & 0.1153 & 0.0893 & -0.0245 & -0.0261 & Rhythmic-favoured \\
\bottomrule
\end{tabular}

\end{table*}

\subsection{Cross-Network Behavioral Transferability}
\label{subsec:transferability}

Before quantifying behavioral information leakage, it is important to
determine whether behavioral signatures transfer across different
anonymity-network environments. Consistent transfer would indicate
that the underlying behavioral patterns generalize beyond a particular
network architecture, whereas poor transfer would demonstrate strong
network dependence.

\begin{figure}[!htbp]
    \centering
    \includegraphics[
        width=0.96\columnwidth,
        trim=5 5 5 5,
        clip
    ]{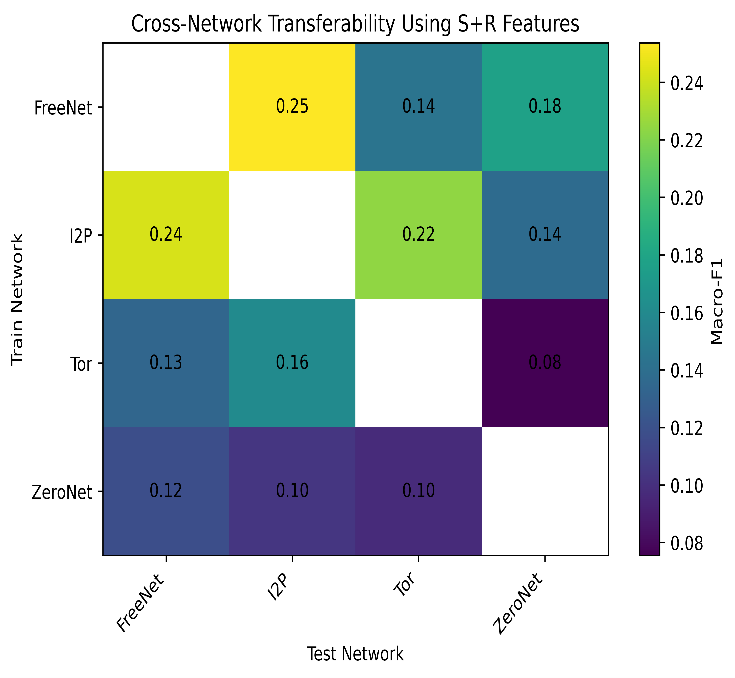}
    \caption{Pairwise cross-network transferability using the combined structural--rhythmic  representation. Rows denote training networks, columns denote test networks, and cell values report Macro-F1.}
    \label{fig:lono}
\end{figure}

Figure~\ref{fig:lono} presents pairwise cross-network
transferability using the combined structural--rhythmic
representation. Each model was trained using traffic from one source
network and evaluated on a different target network. The resulting
Macro-F1 scores remain low across most source--target combinations,
indicating that behavioral patterns learned within one anonymity
network do not transfer reliably to another. Although comparatively
stronger transfer is observed for selected I2P and FreeNet
combinations, the results remain substantially below the corresponding
intra-network performance.

Table~\ref{tab:lono} presents the complementary
leave-one-network-out evaluation. In this experiment, each held-out
network was excluded from training, while the remaining three networks
were combined to train the model. The held-out network was then used
exclusively for testing. Performance remains low across all descriptor
groups. The highest Structural Macro-F1 is obtained when I2P is held
out (\(0.1716\)), whereas the lowest is observed for ZeroNet
(\(0.0908\)). These results demonstrate that even pooled training
across multiple anonymity infrastructures provides limited
generalization to a previously unseen network.

A relative channel difference is also observed in the
leave-one-network-out results. Structural and Rhythmic descriptors
produce similar performance for FreeNet and I2P, although the absolute
Macro-F1 values remain low. For Tor and ZeroNet, the Rhythmic group
outperforms the Structural group by \(0.0600\) and \(0.0245\),
respectively. These differences indicate that temporal descriptors
retain comparatively more transferable information for these held-out
networks; however, they do not imply strong cross-network
generalization.

Another important finding is that combining structural and rhythmic descriptors does not improve cross-network transferability. Across all held-out networks, the combined structural--rhythmic representation (\(S+R\)) performs similarly to, or slightly worse than, the strongest individual behavioral channel. This behavior differs markedly from the within-network experiments presented later in this section, where descriptor fusion consistently improves predictive performance. The result suggests that the additional information captured by descriptor fusion is largely network specific rather than universally transferable. In other words, interactions between the structural and rhythmic behavioral channels appear to encode communication characteristics unique to individual anonymity network architectures.

Overall, the transferability analysis demonstrates that behavioral leakage is not governed by a universal cross-network signature. Instead, encrypted traffic behavior is shaped by interactions between service semantics and the architecture of the anonymity network. Although certain behavioral properties remain partially transferable, the substantial reduction in predictive performance under cross-network evaluation indicates that different anonymity networks generate distinct behavioral ecosystems. This finding motivated the subsequent descriptor-level analysis, which investigates the specific behavioral variables responsible for these network-dependent leakage patterns.
\begin{figure*}[!t]
    \centering
    \includegraphics[
        width=0.88\textwidth,
        trim=3 3 3 3,
        clip
    ]{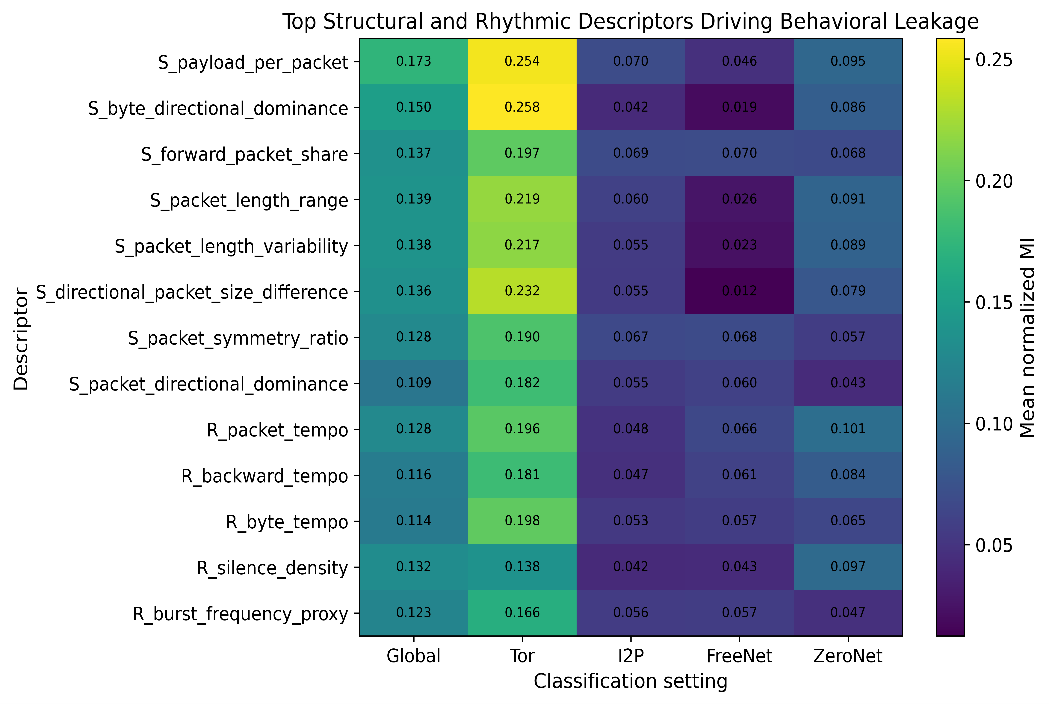}
    \caption{Descriptor-level normalized mutual information across the global and network-specific classification settings.}
    \label{fig:descriptor_mi}
\end{figure*}
\begin{table*}[!htbp]
\centering
\caption{Dominant Structural and Rhythmic descriptors based on normalized mutual information.}
\label{tab:descriptor_mi}

\small
\setlength{\tabcolsep}{10pt}
\renewcommand{\arraystretch}{1.2}

\begin{tabular}{llclc}
\toprule
\textbf{Setting} &
\textbf{Strongest Structural Descriptor} &
\textbf{Normalized MI} &
\textbf{Strongest Rhythmic Descriptor} &
\textbf{Normalized MI} \\
\midrule
Global  & \texttt{S\_payload\_per\_packet}          & 0.173 & \texttt{R\_silence\_density}        & 0.132 \\
Tor     & \texttt{S\_byte\_directional\_dominance} & 0.258 & \texttt{R\_byte\_tempo}             & 0.198 \\
I2P     & \texttt{S\_forward\_packet\_share}       & 0.069 & \texttt{R\_burst\_frequency\_proxy} & 0.056 \\
FreeNet & \texttt{S\_forward\_packet\_share}       & 0.070 & \texttt{R\_packet\_tempo}           & 0.066 \\
ZeroNet & \texttt{S\_payload\_per\_packet}          & 0.095 & \texttt{R\_packet\_tempo}           & 0.101 \\
\bottomrule
\end{tabular}
\end{table*}
\subsection{Descriptor-Level Behavioral Leakage Characterization}
\label{subsec:descriptor-leakage}

Having established that behavioral signatures have limited cross-network transferability, the next step was to identify the specific descriptors responsible for behavioral information leakage. Although the transferability analysis revealed limited generalization
across anonymity networks, it did not identify which observable traffic
characteristics were responsible for the network-dependent behavioral
signatures. Figure~\ref{fig:descriptor_mi} and Table~\ref{tab:descriptor_mi} address this question by examining descriptor-level normalized mutual information across all experimental settings.

Figure~\ref{fig:descriptor_mi} shows that behavioral information leakage is not uniformly distributed across descriptors. Rather, a relatively small subset of structural and rhythmic descriptors consistently contributes a disproportionate share of the total behavioral leakage. This concentration of information indicates that darknet service identifiability is driven by a limited number of communication behaviors rather than by the complete feature space. Consequently, encrypted traffic remains distinguishable, not because every observable property leaks information but because certain behavioral characteristics persist despite encryption and anonymity protection.

The descriptor rankings presented in Table~\ref{tab:descriptor_mi} reveal clear differences among anonymity networks. At the global level, the most informative Structural descriptor is \textit{S\_payload\_per\_packet}, while \textit{R\_silence\_density} emerges as the strongest rhythmic descriptor. This combination indicates that both communication organization and communication pacing contribute substantially to service identifiability when traffic from all anonymity networks is analyzed simultaneously. The presence of highly informative descriptors from both behavioral channels further supports the hypothesis that behavioral leakage is inherently multidimensional.

Among all the anonymity networks, Tor exhibits the strongest descriptor-level behavioral leakage. The dominant structural descriptor, \textit{S\_byte\_directional\_dominance}, achieves a normalized mutual information score of \(0.258\), substantially exceeding the strongest rhythmic descriptor, \textit{R\_byte\_tempo} (\(0.198\)). This result suggests that Tor services are primarily distinguished through directional communication asymmetry rather than temporal behavior alone. Since different Tor applications display varying degrees of bidirectional interaction, directional byte-exchange patterns constitute a particularly informative source of behavioral leakage. This observation is consistent with the structural dominance analysis presented later in Section~\ref{subsec:interaction-analysis}.

By contrast, FreeNet and ZeroNet appear to rely more heavily on rhythmic communication. In both anonymity networks, timing-oriented descriptors such as \textit{R\_packet\_tempo} rank among the most informative variables, indicating that communication pacing and temporal regularity contribute more to service discrimination than the characteristics of packet structures. This finding suggests that the behavioral identity of these networks is encoded primarily through temporal communication dynamics and is consistent with the rhythmic dominance patterns discussed in Section~\ref{subsec:interaction-analysis}.

Of the remaining anonymity networks, I2P shows comparatively weaker descriptor-level behavioral leakage, with both structural and rhythmic descriptors producing substantially lower normalized mutual information values. Nevertheless, the highest-ranked descriptors remain interpretable from a behavior point of view, demonstrating that even modest leakages originate from meaningful communication characteristics rather than statistical artifacts. These findings suggest that I2P services are distinguishable through a combination of packet-sharing behavior and burst dynamics, although the overall leakage remains weaker than that observed for Tor.

A notable observation across all experimental settings is that the highest-ranked descriptors consistently correspond to interpretable behavioral properties, including payload organization, directional communication balance, temporal regularity, and silence behavior. No individual descriptor dominates across every anonymity network, reinforcing the conclusion that behavioral leakage mechanisms depend strongly on the underlying anonymity-network architecture. Consequently, darknet service identifiability cannot be attributed to a universal traffic signature but instead emerges from network-specific combinations of structural and rhythmic communication behavior.

Overall, the descriptor-level analysis provides direct evidence that behavioral information leakage originates from a relatively small number of highly informative communication characteristics. These findings establish the behavioral foundation for the subsequent channel-level leakage analysis, where the cumulative contributions of the structural and rhythmic descriptor groups are quantified.

\begin{table*}[!hbtp]
\centering
\caption{Network-level predictive performance and cumulative behavioral leakage using Random Forest over 50 repeated stratified cross-validation splits. Here, \(F_1\) denotes Macro-F1, \(L\) denotes cumulative normalized leakage, and Gain denotes the improvement of \(S+R\) over the strongest individual behavioral group.}
\label{tab:network_leakage}

\small
\setlength{\tabcolsep}{6pt}
\renewcommand{\arraystretch}{1.2}

\begin{tabular}{lccccccccc}
\toprule
\textbf{Setting} &
\(\boldsymbol{F_1^C}\) &
\(\boldsymbol{F_1^S}\) &
\(\boldsymbol{F_1^R}\) &
\(\boldsymbol{F_1^{S+R}}\) &
\(\boldsymbol{L_C}\) &
\(\boldsymbol{L_S}\) &
\(\boldsymbol{L_R}\) &
\(\boldsymbol{L_{S+R}}\) &
\textbf{Gain (\%)} \\
\midrule

Global &
\(0.3348 \pm 0.0056\) &
\(0.4388 \pm 0.0058\) &
\(0.4560 \pm 0.0063\) &
\(0.5251 \pm 0.0059\) &
0.5936 &
1.3604 &
1.3654 &
2.7252 &
15.15 \\

Tor &
\(0.6002 \pm 0.0115\) &
\(0.6642 \pm 0.0105\) &
\(0.6683 \pm 0.0112\) &
\(0.7165 \pm 0.0105\) &
0.6941 &
2.1760 &
1.7624 &
3.9461 &
7.21 \\

I2P &
\(0.3778 \pm 0.0084\) &
\(0.4352 \pm 0.0114\) &
\(0.4312 \pm 0.0143\) &
\(0.4855 \pm 0.0144\) &
0.2314 &
0.5393 &
0.4635 &
1.0001 &
11.53 \\

FreeNet &
\(0.2458 \pm 0.0069\) &
\(0.3731 \pm 0.0081\) &
\(0.4111 \pm 0.0138\) &
\(0.4510 \pm 0.0085\) &
0.0615 &
0.4156 &
0.4614 &
0.8744 &
9.71 \\

ZeroNet &
\(0.3631 \pm 0.0081\) &
\(0.3773 \pm 0.0102\) &
\(0.4762 \pm 0.0125\) &
\(0.5162 \pm 0.0112\) &
0.3636 &
0.7724 &
0.8014 &
1.5717 &
8.40 \\

\bottomrule
\end{tabular}
\end{table*}
\begin{figure*}[!t]
    \centering
    \includegraphics[
        width=1\textwidth,
        trim=3 3 3 3,
        clip
    ]{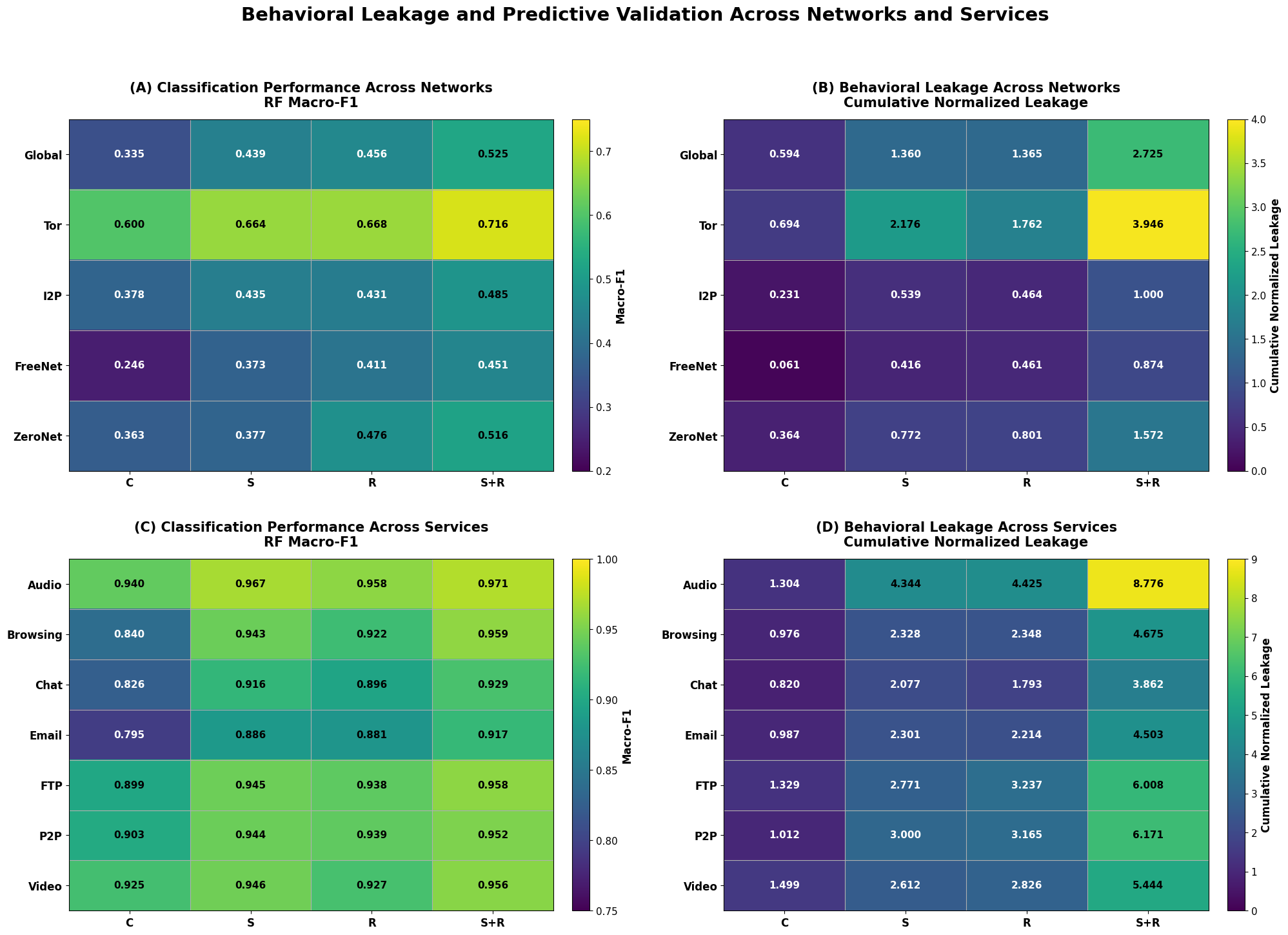}
    \caption{Predictive validation and cumulative behavioral leakage across network-level and semantic-service classification settings. Panels (A) and (C) report RF Macro-F1, while Panels (B) and (D) report cumulative normalized leakage for the corresponding descriptor representations. Values represent means across the evaluated repeated cross-validation splits or available semantic network pairs.}
    \label{fig:leakage_validation}
\end{figure*}
\subsection{Behavioral Leakage Quantification and Predictive Validation}
\label{subsec:leakage-validation}

A fundamental question in behavioral leakage analysis is whether the information exposed by encrypted traffic is sufficiently discriminative to reliably identify services. Although the descriptor-level analysis demonstrates that individual structural and rhythmic descriptors contain measurable behavioral leakage, the practical significance of this information depends on whether it can be effectively exploited by a machine learning model. Consequently, both information-theoretic leakage and predictive performance must be evaluated simultaneously.

Figure~\ref{fig:leakage_validation} illustrates the relationship between predictive performance and behavioral information leakage across all descriptor groups.

In all experimental settings, the behavioral descriptor groups consistently expose substantially more information than the control descriptor group. We observed this trend at both the network and service levels, indicating that observable communication behavior contributes more to service identifiability than protocol-level metadata alone. More importantly, increased leakages in behavioral information are consistently accompanied by corresponding improvements in predictive performance, demonstrating that the exposed information represents an exploitable behavioral structure rather than merely statistical noise.

At the global level, both the structural and rhythmic descriptor groups individually improved the Macro-F1 score by more than \(30\%\) relative to the control baseline, while simultaneously more than doubling the normalized cumulative behavioral leakage. Moreover, we observed similar behavior across all the anonymity networks. Tor demonstrated the strongest leakage, with a combined structural--rhythmic representation (\(S+R\))  of \(3.9461\) in terms of normalized cumulative leakage, together with a Macro-F1 score of \(0.7165\). By contrast, FreeNet produced the lowest combined cumulative leakage (\(0.8744\)) and the lowest intra-network Macro-F1 (\(0.4510\)), suggesting comparatively weaker service-level behavioral separability in the evaluated descriptor space.

The combined structural--rhythmic representation achieves the highest
predictive performance across all global and intra-network settings.
The positive Macro-F1 gain beyond the stronger individual channel
provides direct evidence that Structural and Rhythmic descriptors
contain complementary predictive information. The combined
representation also produces the largest cumulative leakage; however,
this quantity is interpreted cautiously because it aggregates marginal
descriptor-level contributions and is partly influenced by the larger
number of retained descriptors.

A broad correspondence is observed between cumulative behavioral leakage and predictive performance. Networks with higher cumulative leakage generally achieve stronger predictive separability, whereas
lower leakage is associated with weaker classification performance. This correspondence supports the interpretation that the quantified
leakage reflects behavioral information that can be exploited for encrypted service identification.

The complete network-level and service-level leakage distributions are illustrated in Figure~\ref{fig:leakage_validation}, while the corresponding quantitative results are summarized in Table~\ref{tab:network_leakage}.

The relatively small standard deviations across the 50 repeated stratified splits indicate limited sensitivity to the evaluated random partitions. It should also be noted that the leakage values reported in Table~\ref{tab:network_leakage} represent cumulative normalized behavioral leakage obtained by summing descriptor-level mutual information contributions after normalizing the values according to label entropy. Consequently, these values quantify aggregate behavioral information exposure across descriptor groups and are therefore not constrained to the \([0,1]\) interval associated with normalized mutual information for an individual descriptor.

An interesting observation is that the global classification setting benefits most from descriptor fusion. This indicates that multi-network environments require both structural and rhythmic behavioral information to reliably discriminate between services. Tor exhibits a comparatively smaller fusion gain because both the Structural and Rhythmic representations already achieve strong and
nearly balanced single-channel performance, leaving less additional predictive improvement available to the combined representation.

\subsection{Behavioral Mechanisms and Structural--Rhythmic Interactions}
\label{subsec:behavioral-mechanisms}

Behavioral information leakage originates from multiple observable characteristics embedded within encrypted communication flows. Although individual descriptors provide measurable information regarding service identity, they frequently represent broader communication mechanisms that interact across multiple behavioral channels. Consequently, understanding service identifiability requires not only quantifying behavioral leakage but also investigating the underlying communication mechanisms responsible for information exposure and the interactions between the structural and rhythmic behavioral channels.

Accordingly, this section examines behavioral leakage from two complementary perspectives. First, descriptor-level information is aggregated into higher-level communication mechanisms to identify the dominant sources of information exposure. Second, interactions between the structural and rhythmic descriptor groups are investigated to determine whether they provide redundant or complementary behavioral information.

\subsubsection{Mechanism-Level Behavioral Leakage Composition}
\label{subsec:mechanism-composition}

Individual descriptors frequently capture related aspects of communication behavior and collectively contribute to higher-level behavioral leakage mechanisms. To obtain a more interpretable view of information exposure, we aggregated descriptor-level contributions into five communication mechanisms:

\begin{itemize}[leftmargin=*]

\item Packet-size and payload-length structure;

\item Directionality and exchange balance;

\item Timing rhythm;

\item Silence and burst behavior; and

\item Residual descriptors.

\end{itemize}

Here, payload structure refers exclusively to observable packet-size and payload-length statistics and does not involve inspection of encrypted packet contents.

For each behavioral mechanism \(m\), its cumulative normalized leakage
was calculated as

\begin{equation}
L_m
=
\frac{1}{H(Y)}
\sum_{f_j \in \mathcal{F}_m}
I(f_j;Y),
\label{eq:mechanism_leakage}
\end{equation}

where \(\mathcal{F}_m\) denotes the set of descriptors assigned to mechanism \(m\). Thus, the mechanism-level values represent the aggregate normalized marginal mutual-information contributions of all descriptors mapped to the corresponding communication mechanism.

This aggregation provided a mechanism-oriented interpretation of service identifiability that extended beyond the importance of individual descriptors.

\begin{figure*}[!htbp]
    \centering
    \includegraphics[
        width=0.75\textwidth,
        trim=3 3 3 3,
        clip
    ]{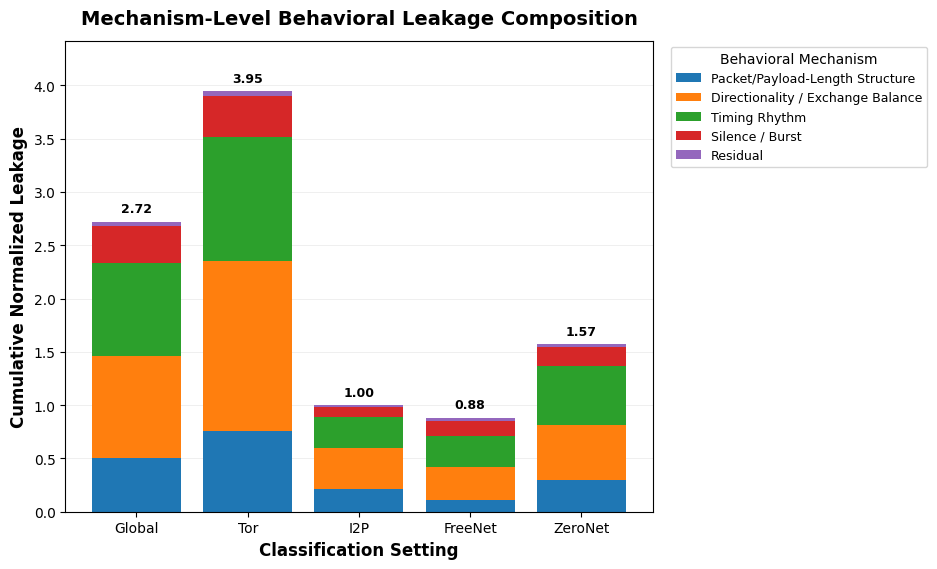}
    \caption{
Mechanism-level decomposition of behavioral information leakage.}
    \label{fig:mechanism_leakage}
\end{figure*}

The mechanism-level results presented in
Figure~\ref{fig:mechanism_leakage} and
Table~\ref{tab:mechanism_leakage} show that behavioral information
leakage is distributed across multiple communication mechanisms rather
than being concentrated within a single traffic characteristic.
Directionality and exchange balance provide the largest cumulative
contribution in the Global, Tor, I2P, and FreeNet settings, whereas
Timing Rhythm contributes slightly more than directionality for
ZeroNet. Across the evaluated settings, these two mechanisms therefore
represent the principal sources of aggregate behavioral leakage.

Packet-size and payload-length structure provides an additional but
smaller contribution, while silence and burst behavior contributes
comparatively less information. Residual descriptors make only a minor
contribution in every setting. These findings indicate that encrypted
service distinguishability is primarily shaped by the joint effects of
directional exchange organization and temporal communication rhythm
rather than by a single universally dominant mechanism. The complete
numerical contributions are reported in
Table~\ref{tab:mechanism_leakage}, while their relative composition is
visualized in Figure~\ref{fig:mechanism_leakage}.

Temporal communication mechanisms also contribute substantial behavioral information. Timing rhythm captures packet tempo and communication periodicity, whereas silence and burst descriptors characterize inactivity intervals and communication fluctuations. Although these temporal mechanisms generally contribute less information than directional communication behavior, they consistently provide additional discriminatory information across every anonymity network.

Distinct mechanism-level profiles did emerge among the evaluated anonymity networks. For example, Tor exhibited the strongest contributions from directionality and exchange balance together with timing rhythm, indicating that onion-routing traffic preserves highly characteristic communication structures and temporal regularities. By contrast, I2P exhibited comparatively smaller contributions across all communication mechanisms, suggesting greater overlap among service behaviors. FreeNet and ZeroNet derive a relatively larger proportion of their behavioral leakage from temporal communication mechanisms, consistent with the descriptor-level observations presented in Section~\ref{subsec:descriptor-leakage}, where packet-tempo and silence-density descriptors emerged as being among the most informative behavioral variables.

These observations demonstrate that service identifiability arises through the combined influence of packet organization, directional communication patterns, temporal rhythm, and burst dynamics. Consequently, behavioral information leakage should be interpreted as a mechanism-level phenomenon rather than as the consequence of isolated descriptors. The complete mechanism-level decomposition is illustrated in Figure~\ref{fig:mechanism_leakage}, while the corresponding cumulative leakage contributions are summarized in Table~\ref{tab:mechanism_leakage}.

\begin{table*}[!t]
\centering
\caption{Mechanism-level behavioral leakage composition based on cumulative normalized mutual-information contributions.}
\label{tab:mechanism_leakage}
\small
\setlength{\tabcolsep}{12pt}
\renewcommand{\arraystretch}{1.5}

\begin{tabular}{lcccccc}
\toprule
\textbf{Setting}
&
{\makecell{\textbf{Packet/Payload-Length}\\\textbf{Structure}}}
&
{\makecell{\textbf{Directionality/}\\\textbf{Exchange Balance}}}
&
{\makecell{\textbf{Timing}\\\textbf{Rhythm}}}
&
{\makecell{\textbf{Silence/}\\\textbf{Burst}}}
&
{\textbf{Residual}}

\\
\midrule
Global  & 0.501 & 0.958 & 0.873 & 0.353 & 0.037 \\
Tor     & 0.756 & 1.592 & 1.165 & 0.389 & 0.044 \\
I2P     & 0.215 & 0.382 & 0.296 & 0.089 & 0.018 \\
FreeNet & 0.114 & 0.306 & 0.295 & 0.140 & 0.029 \\
ZeroNet & 0.301 & 0.516 & 0.546 & 0.184 & 0.025 \\
\bottomrule
\end{tabular}
\end{table*}

\begin{figure*}[!t]
    \centering
    \includegraphics[
        width=1\textwidth,
        trim=3 3 3 3,
        clip
    ]{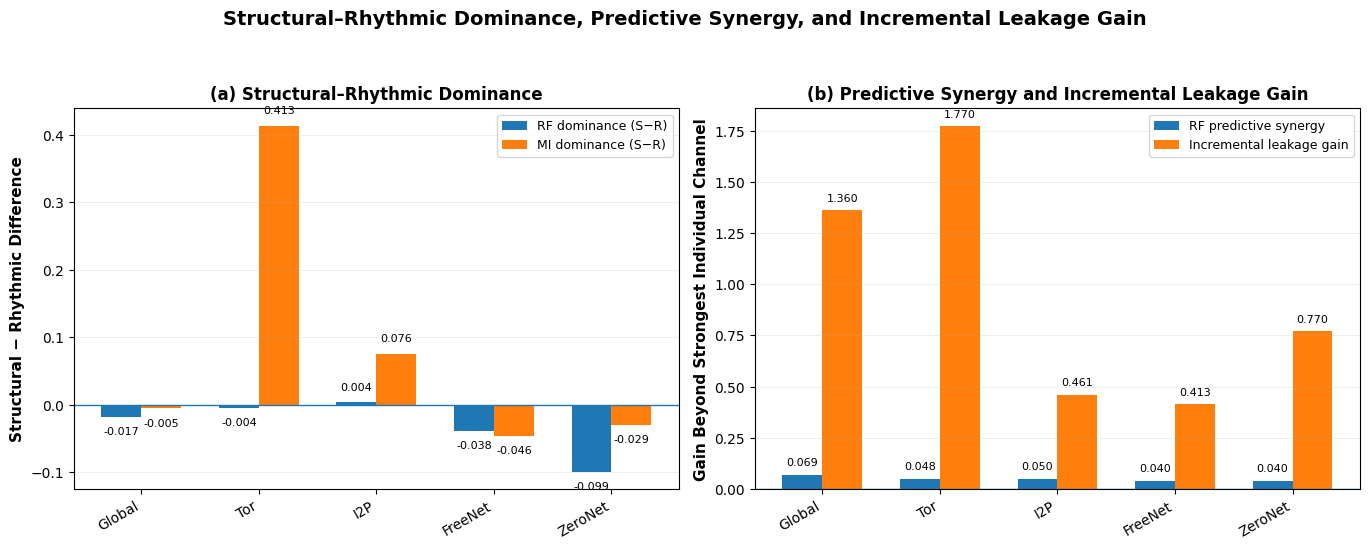}
    \caption{Structural--rhythmic dominance
(\(D_{\mathrm{RF}},D_{\mathrm{MI}}\)), predictive synergy
(\(\Delta_{\mathrm{RF}}\)), and incremental cumulative leakage gain
(\(\Delta_L\)) across the global and network-specific experimental
settings.}
    \label{fig:synergy}
\end{figure*}
\subsubsection{Structural--Rhythmic Interaction Analysis}
\label{subsec:interaction-analysis}

Figure~\ref{fig:synergy} and
Table~\ref{tab:synergy} summarize the relative dominance of
the Structural and Rhythmic descriptor groups, their predictive
synergy, and the incremental cumulative leakage exposed by their
combination.

\begin{table*}[!hbtp]
\centering
\caption{Structural--rhythmic dominance, predictive synergy, and
incremental cumulative leakage gain across the global and
network-specific settings.}
\label{tab:synergy}
\small
\setlength{\tabcolsep}{14pt}
\renewcommand{\arraystretch}{1.2}

\begin{tabular}{
l
S[table-format=-1.4]
S[table-format=-1.4]
S[table-format=1.4]
S[table-format=1.4]
}
\toprule
\textbf{Setting}
&
{\boldmath$\mathbf{D}_{\mathrm{RF}}$}
&
{\boldmath$\mathbf{D}_{\mathrm{MI}}$}
&
{\boldmath$\boldsymbol{\Delta}_{\mathrm{RF}}$}
&
{\boldmath$\boldsymbol{\Delta}_{L}$}
\\
\midrule
Global  & -0.0172 & -0.0049 & 0.0691 & 1.3598 \\
Tor     & -0.0042 &  0.4135 & 0.0481 & 1.7701 \\
I2P     &  0.0040 &  0.0757 & 0.0502 & 0.4609 \\
FreeNet & -0.0380 & -0.0458 & 0.0399 & 0.4130 \\
ZeroNet & -0.0988 & -0.0290 & 0.0400 & 0.7703 \\
\bottomrule
\end{tabular}

\begin{flushleft}
\footnotesize
Positive dominance values indicate Structural dominance, whereas
negative values indicate Rhythmic dominance. Here,
\(\Delta_{\mathrm{RF}}\) denotes predictive synergy and
\(\Delta_L\) denotes incremental cumulative leakage gain.
\end{flushleft}
\end{table*}

Positive predictive synergy is observed across the global and all
intra-network settings. The combined structural--rhythmic
representation improves Macro-F1 beyond the stronger individual
descriptor group by \(0.0691\) globally, \(0.0481\) for Tor,
\(0.0502\) for I2P, \(0.0399\) for FreeNet, and \(0.0400\) for
ZeroNet. These improvements demonstrate that the two channels contain
complementary information that can be jointly exploited for service
classification.

The incremental cumulative leakage gain also remains positive across
all settings. The global setting records an additional aggregate
leakage contribution of \(1.3598\), while Tor exhibits the largest
network-specific gain of \(1.7701\). I2P, FreeNet, and ZeroNet produce
smaller gains of \(0.4609\), \(0.4130\), and \(0.7703\),
respectively. These values indicate that adding the second behavioral
channel contributes further descriptor-level leakage beyond the
stronger individual group. Because cumulative leakage is based on
summed marginal mutual-information contributions, these gains are not
interpreted as formal information-theoretic synergy.

The dominance results further reveal architecture-dependent
differences. Tor exhibits strong Structural dominance in cumulative
leakage (\(D_{\mathrm{MI}}=0.4135\)), although its predictive
dominance is close to zero
(\(D_{\mathrm{RF}}=-0.0042\)). I2P also exhibits moderate Structural
dominance in cumulative leakage while remaining predictively balanced.
By contrast, FreeNet and ZeroNet exhibit Rhythmic dominance from both
predictive and leakage perspectives.

Overall, Structural descriptors characterize packet-size
organization, directional asymmetry, and exchange balance, whereas
Rhythmic descriptors characterize packet tempo, burst activity,
silence intervals, and timing regularity. Their positive predictive
synergy demonstrates that encrypted service identifiability arises
from complementary structural and temporal manifestations of
communication behavior. Figure~\ref{fig:synergy} provides
the graphical comparison, while Table~\ref{tab:synergy}
reports the corresponding numerical results.

\subsection{Service-Level Variability of Cross-Network Separability}
\begin{table*}[!htbp]
\centering
\caption{Pairwise network-origin Macro-F1 and cross-network service-level variability. FN, T, and ZN denote FreeNet, Tor, and ZeroNet, respectively. A dash indicates that the corresponding semantic comparison was unavailable.}
\label{tab:service_synergy}

\small
\setlength{\tabcolsep}{14pt}
\renewcommand{\arraystretch}{1.2}

\begin{tabular}{lcccccccc}
\toprule
\textbf{Service} &
\textbf{FN--I2P} &
\textbf{FN--T} &
\textbf{FN--ZN} &
\textbf{I2P--T} &
\textbf{I2P--ZN} &
\textbf{T--ZN} &
\textbf{SVI} &
\textbf{LVI} \\
\midrule
Audio    & --     & --     & --     & --     & --     & 0.9708 & --     & --     \\
Video    & --     & 0.9579 & 0.9595 & --     & --     & 0.9503 & 0.0049 & 0.2961 \\
FTP      & 0.9655 & 0.9441 & 0.9703 & 0.9553 & 0.9637 & 0.9470 & 0.0106 & 1.6864 \\
P2P      & --     & --     & --     & 0.9325 & 0.9716 & 0.9513 & 0.0195 & 1.0363 \\
Browsing & 0.9470 & 0.9472 & 0.9825 & 0.9549 & 0.9647 & 0.9557 & 0.0134 & 1.6232 \\
Chat     & 0.9332 & 0.9459 & 0.9482 & 0.8843 & 0.9338 & 0.9256 & 0.0232 & 1.0063 \\
Email    & 0.9461 & 0.9230 & 0.9366 & 0.8596 & 0.9197 & 0.9190 & 0.0303 & 1.7353 \\
\bottomrule
\end{tabular}
\end{table*}

\begin{figure*}[!t]
    \centering
    \includegraphics[
        width=1\textwidth,
        trim=3 3 3 3,
        clip
    ]{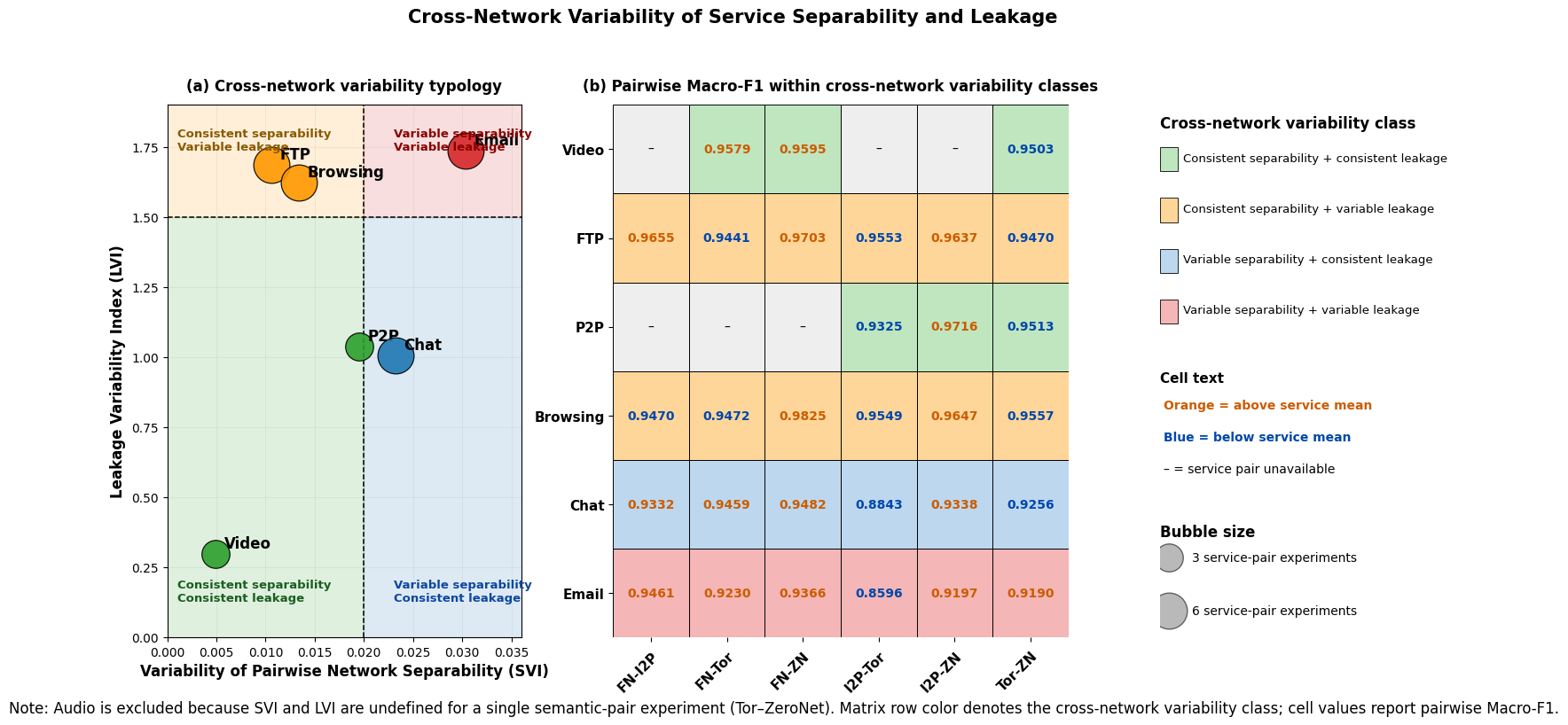}
   \caption{Cross-network variability of service-level network separability and behavioral leakage: (a) services positioned according to the Service Variability Index (SVI) and Leakage Variability Index
(LVI), and (b) pairwise network-origin Macro-F1 scores for each semantic service.}
\label{fig:service_variability}
\end{figure*}

The previous analyses demonstrate that Structural and Rhythmic
behavioral characteristics vary across anonymity networks. The
semantic-pair experiments further examine whether the degree to which
a fixed service reveals its network origin remains consistent across
different anonymity-network combinations.

For each service, pairwise classifiers distinguish the anonymity network associated with otherwise equivalent semantic traffic using the combined structural--rhythmic representation. The resulting Macro-F1 values therefore quantify network-specific separability rather than service-label transferability. SVI measures the variation of these pairwise Macro-F1 values, while LVI measures variation in their cumulative normalized leakage. Low SVI indicates consistent network-specific separability across pairs, whereas low LVI indicates consistent leakage magnitude.

Video exhibits the most consistent network-specific separability
across the available semantic pairs. Its pairwise Macro-F1 values range from \(0.9503\) to \(0.9595\), producing an SVI of \(0.0049\). Its LVI of \(0.2961\) further indicates comparatively consistent leakage magnitude. These results show that architecture-dependent differences in video traffic are expressed
consistently across Tor, FreeNet, and ZeroNet; they do not imply that video has an architecture-invariant traffic signature.

Browsing and FTP also exhibit relatively consistent network-specific
separability, with SVI values of \(0.0134\) and \(0.0106\),
respectively. Their larger LVI values of \(1.6232\) and \(1.6864\)
show that, although their pairwise classification performance remains
comparatively consistent, the amount of exposed behavioral information
varies substantially among network combinations. Thus, stable
pairwise Macro-F1 does not necessarily imply stable leakage magnitude.

P2P occupies an intermediate position, with an SVI of \(0.0195\)
and an LVI of \(1.0363\). Its network-specific separability therefore
varies more than that of video, FTP, and browsing, while its leakage
magnitude shows moderate variation. This pattern is consistent with
the combination of persistent data exchange, peer discovery, overlay
management, and connection establishment that characterizes
peer-to-peer communication.

Chat and email exhibit the greatest variation in network-specific
separability. Email has the largest SVI (\(0.0303\)) and LVI
(\(1.7353\)), while chat has an SVI of \(0.0232\) and an LVI of
\(1.0063\). The lowest pairwise Macro-F1 occurs for email in the
I2P--Tor comparison (\(0.8596\)). These results suggest that interactive and asynchronous communication
services produce more variable architecture-dependent signatures than continuously operating protocol-driven services.

The pairwise semantic results summarized in
Table~\ref{tab:service_synergy} further illustrate how individual
network combinations contribute to these cross-network variability
profiles. Rather than exhibiting uniform behavior across all semantic
network pairs, several services demonstrate sensitivity to specific
combinations of anonymity networks. Browsing exhibits the strongest
network-origin separability in the FreeNet--ZeroNet comparison
(\(0.9825\)), whereas email exhibits the lowest network-origin
separability in the I2P--Tor comparison (\(0.8596\)). These
observations indicate that anonymity-network implementations are
associated with different Structural and Rhythmic behavioral profiles,
producing varying levels of network-origin separability for otherwise
equivalent semantic services.

To provide a unified interpretation of these observations,
Figure~\ref{fig:service_variability} summarizes the variability of
network-specific service separability and leakage magnitude.
Figure~\ref{fig:service_variability}(a) positions each service
according to its SVI and LVI values. Under the descriptive operational
cut-offs used in the visualization, video and P2P exhibit comparatively
consistent pairwise separability and leakage magnitude, while browsing
and FTP combine consistent separability with more variable leakage.
Chat exhibits variable separability with comparatively consistent
leakage, whereas email exhibits variability in both measures.

Figure~\ref{fig:service_variability}(b) presents the corresponding
pairwise network-origin Macro-F1 scores. The matrix identifies the
network combinations for which the same semantic service reveals
stronger or weaker architecture-dependent differences.

Overall, the semantic-pair analysis demonstrates that the consistency of network-specific separability depends jointly on service semantics and anonymity-network architecture. Protocol-driven services such as video exhibit consistently high network-origin separability, whereas interactive services such as chat and email show greater variation across network pairs. SVI and LVI therefore haracterize variation in pairwise network separability and leakage magnitude rather than architecture-invariant service signatures.

\section{Discussion}

The proposed behavioral decomposition framework provides a complementary perspective to conventional encrypted traffic classification by explaining \emph{why} darknet services remain distinguishable despite encryption rather than simply examining \emph{how accurately} they can be classified. Integrating information-theoretic leakage estimation with predictive validation demonstrates that encrypted service identifiability is governed by multiple observable behavioral channels rather than by
isolated statistical descriptors~\cite{ENISA2020,velan2015survey,Montieri2020Dive}.

The results show that structural and rhythmic descriptors capture different aspects of encrypted communication behavior~\cite{panchenko2016website,Hayes2016KFingerprinting}. Structural descriptors characterize packet organization, directional communication, and traffic exchange patterns, whereas rhythmic descriptors describe temporal dynamics through packet tempo, burst behavior, and silence intervals~\cite{Sirinam2018DeepFingerprinting}. Their consistently positive within-network synergy indicates that these channels encode complementary behavioral information, explaining why the combined structural--rhythmic representation achieves superior predictive performance and cumulative leakage in the global and
intra-network experimental settings.

Behavioral leakage is also strongly influenced by
anonymity-network architecture. Tor exhibits strong Structural dominance in cumulative leakage, while its predictive performance remains nearly balanced between the Structural and Rhythmic channels. FreeNet and ZeroNet, by contrast, exhibit Rhythmic dominance from both predictive and cumulative-leakage perspectives. I2P, meanwhile, presents weaker and more balanced leakage, suggesting greater overlap among service behaviors. These observations indicate that anonymity-network architectures are
associated with differences in both the magnitude of observable information and the behavioral channels through which it is exposed.~\cite{Montieri2020Dive, Karunanayake2023ModifiedTor, saleem2024darknet}.

The proposed Service Variability Index (SVI) and Leakage Variability
Index (LVI) extend conventional evaluation by quantifying variation in
network-specific service separability and leakage magnitude across
semantic network pairs. Video exhibits consistently high
network-origin separability and comparatively stable leakage, whereas
chat and email show greater variation among network combinations.
These findings indicate that architecture-dependent behavioral
differences are expressed differently across service categories and
are jointly influenced by application semantics and
anonymity-network design.

From a practical perspective, the proposed framework supports interpretable traffic analysis, privacy auditing, and the evaluation of traffic-obfuscation strategies. Since packet-size organization and temporal behavior contribute complementary information, effective privacy-preserving defenses should jointly address multiple behavioral channels instead of simply addressing packet size or timing independently~\cite{Juarez2016WTFPAD,Cadena2020TrafficSliver}.

Several limitations should be considered when interpreting these findings. First, the evaluation uses a single publicly available flow-level dataset, and the observed leakage profiles may vary across capture environments, software versions, and traffic-collection periods. Second, Random Forest is used as a predictive validation model rather than as an exhaustive comparison across classifier families. Third, cumulative normalized leakage aggregates marginal descriptor-level mutual-information contributions and may therefore contain overlapping information. Finally, the number of available semantic network pairs differs among services, with audio represented by only one pair. Future cross-dataset and session-aware evaluation is therefore required before broader operational generalization.

\section{Conclusion}

This paper presented a behavioral decomposition framework for quantifying and interpreting information leakage in encrypted darknet traffic across multiple anonymity networks. By integrating information-theoretic leakage quantification with predictive validation, the proposed framework decomposes encrypted flow descriptors into control, structural, rhythmic, and combined structural--rhythmic representations. Hence, we were able to analyze behavioral leakage without needing to inspect the contents of the payload.

Experimental evaluation on the publicly available Darknet Dataset 2020 demonstrates that encrypted traffic continues to expose measurable behavioral information despite anonymity protections. The results show that the relative contributions of Structural and
Rhythmic descriptors vary across Tor, I2P, FreeNet, and ZeroNet, while their combination provides the highest predictive performance and cumulative leakage in the global and intra-network settings.
Leave-one-network-out evaluation, however, demonstrates that these combined behavioral signatures remain strongly architecture dependent.

Additionally, the proposed Service Variability Index (SVI) and Leakage
Variability Index (LVI) quantify variation in network-specific service
separability and cumulative leakage across semantic network pairs.
These measures show that architecture-dependent behavioral differences
are expressed consistently for some services, such as video, but vary
more substantially for interactive services such as chat and email.
Together with the dominance, predictive-synergy, and mechanism-level
analyses, these findings demonstrate that encrypted service
identifiability is jointly shaped by communication structure, temporal
dynamics, application semantics, and anonymity-network architecture.

Future work will evaluate the robustness of the proposed framework under adaptive traffic obfuscation, packet padding, timing perturbation, traffic morphing, and traffic splitting, while extending the analysis to additional datasets, emerging anonymity protocols, and cross-dataset generalization scenarios.


\section*{Credit authorship contribution statement}

Javeriah Saleem: conceptualization, methodology, software,
formal analysis, investigation, data curation, visualization,
writing -- original draft.

Rafiqul Islam: supervision, conceptualization, validation,
writing -- review and editing.

Md Zahidul Islam: supervision, methodology, validation,
writing -- review and editing.

\section*{Declaration of competing interest}

The authors declare that they have no known competing financial
interests or personal relationships that could have
influenced the work reported in this paper.

\section*{Funding}

This research did not receive any specific grant from funding agencies
in the public, commercial, or not-for-profit sectors.

\section*{Data availability}

The data used in this study are publicly available. The relevant
sources and access information are provided in the manuscript.

\section*{Acknowledgements}

The authors would like to thank Charles Sturt University for providing
the research facilities and academic support required for this study.


\bibliographystyle{elsarticle-num}
\bibliography{references_JISA_behavioral_leakage}
\end{document}